\documentclass[12pt,letterpaper]{article}
\pdfoutput=1
\usepackage{graphicx}
\usepackage{caption}
\usepackage{subcaption}
\usepackage{amsfonts,amssymb,amsmath}

\newcommand{\be}{\begin{equation}}
\newcommand{\ee}{\end{equation}}
\newcommand{\ben}{\begin{displaymath}}
\newcommand{\een}{\end{displaymath}}
\newcommand{\bea}{\begin{eqnarray}}
\newcommand{\eea}{\end{eqnarray}}
\newcommand{\bean}{\begin{eqnarray*}}
\newcommand{\eean}{\end{eqnarray*}}

\newcommand{\ads}[1]{\mbox{${AdS}_{#1}$}}
\newcommand{\adss}[2]{\mbox{$AdS_{#1}\times {S}^{#2}$}}

\newcommand{\ie}{{\it i.e.}}

\newcommand{\tr}{\mbox{Tr}}

\newcommand{\commentout}[1]{}

\newcommand{\X}{\ensuremath{\mathbb{X}}}
\newcommand{\lp}[1]{\ensuremath{\lambda_{p_{#1}}}}
\newcommand{\vect}[2]{\ensuremath{\left[\begin{array}{c}#1\\ #2\end{array}\right]}}
\newcommand{\pds}{\ensuremath{\partial_\sigma}}
\newcommand{\pdt}{\ensuremath{\partial_\tau}}

\newcommand{\beq}{\begin{equation}}
\newcommand{\eeq}{\end{equation}}
\newcommand{\beqa}{\begin{eqnarray}}
\newcommand{\eeqa}{\end{eqnarray}}
\newcommand{\beqar}{\begin{eqnarray*}}
\newcommand{\eeqar}{\end{eqnarray*}}
\newcommand{\cN}{{\cal N}}

\newcommand{\cA}{{\cal A}}
\newcommand{\cB}{{\cal B}}

\newcommand{\half}{\ensuremath{\frac{1}{2}}}

\newcommand{\re}{\ensuremath{{\mathcal Re}}}
\newcommand{\im}{\ensuremath{{\mathcal Im}}}

\newcommand{\N}[1]{\ensuremath{\cN=#1}}

\newcommand{\cU}{\ensuremath{\mathcal{U}}}

\begin{document}

\title{\LARGE \bf Rotating Wilson loops and open strings in \ads{3}}

\author{Andrew Irrgang, Martin Kruczenski\thanks{E-mail: \texttt{irrgang@purdue.edu,markru@purdue.edu}}\\
        Department of Physics, Purdue University,  \\
        525 Northwestern Avenue, W. Lafayette, IN 47907-2036. }

\maketitle

\begin{abstract}
 The AdS/CFT correspondence relates \N{4} super Yang-Mills on $S^3$ to type IIB string theory on \adss{5}{5}. 
 In this context, a quark/anti-quark pair moving on an $S^1\subset S^3$ following prescribed trajectories is dual to an open string 
 ending on the boundary of \ads{3}. In this paper we study the corresponding classical string solutions.
 The Pohlmeyer reduction reduces the equations of motion to a generalized sinh-Gordon equation. This equation
 includes, as particular cases, the Liouville equation as well as the sinh/cosh-Gordon equations. We study generic solutions of
 the Liouville equation and finite gap solutions of the sinh/cosh-Gordon ones. The latter ones are written in terms of Riemann 
 theta functions. The corresponding string solutions are reconstructed giving new solutions and a broad understanding of open 
 strings moving in \ads{3}. As a further example, the simple case of a rigidly rotating string is shown to exhibit these three 
 behaviors depending on a simple relation between its angular velocity and the angular separation between its endpoints.
\end{abstract}

\clearpage
\newpage



\section{Introduction}\label{sec1}
 
 The AdS/CFT correspondence establishes a duality between certain gauge theories and string theory \cite{malda}. The most well-known, and the one we consider here, is between \N{4} super Yang-Mills on $S^3$ and
 type IIB string theory on \adss{5}{5}. This duality has motivated the study of strings moving in AdS backgrounds. Two kinds of strings play an important role. One are closed strings moving inside \ads{5},  which are dual to single trace operators in the gauge theory. 
They have played a most important role in unraveling the AdS/CFT correspondence. For a review see \cite{Tseytlin:2010jv}. 
 The other are open strings ending in the boundary, which are dual to Wilson loops in the gauge theory \cite{MRY}. 
 The case we consider here is time-like Wilson loops that can be understood physically as an (infinitely) heavy quark/anti-quark pair moving along prescribed trajectories. In this paper, we study the case where
 the motion is on a circle which corresponds to an open string moving in \ads{3}. To solve the equations of motion for the string, we first use the Pohlmeyer reduction \cite{Pohlmeyer} to obtain a generalized 
 sinh-Gordon equation for the world-sheet metric. This equation can be reduced, under certain conditions, to the Liouville, the sinh-Gordon or cosh-Gordon equations. In the case of the Liouville equation, we discuss its most generic solution explicitly which turns out to be equivalent to the solutions obtained by Mikhailov in \cite{Mikhailov:2003er}. In the case of the cosh- and sinh-Gordon, we discuss finite gap solution that can be written in terms of theta functions. This gives a rather generic picture of the type of solutions that appear. The solutions in terms of theta functions are closely related to the solutions for closed strings described by Jevicki and Jin \cite{JJ} (see also \cite{Jevicki:2007aa,spiky}) and, more in particular, by Dorey and Vicedo in  \cite{DV}, Sakai and Satoh in \cite{SS}. Moreover, for the case of Wilson loops, theta functions were recently used to find surfaces dual to an infinite parameter family of closed Wilson loops \cite{Ishizeki:2011bf}. It should be noted that the study of strings moving in AdS space predates the AdS/CFT correspondence for example in the work of de Vega and Sanchez \cite{DeVega:1992xc} (see also \cite{Larsen:1996gn}) where the relation with the sinh/cosh-Gordon equation is discussed. The present paper applies those techniques to the case of open strings.

 In a broader context, the study of Wilson loops has provided deep insight into the AdS/CFT correspondence, for example in the case of the circular Wilson loop \cite{cWL,lens,DGO,DF}, the light-like cusp as applied to computation of anomalous dimensions of twist two operators \cite{cusp}, or more recently to scattering amplitudes \cite{AM, AM2, scatampl}. The Pohlmeyer reduction as used here is well-known \cite{Pohlmeyer} but further applications of the method are still being developed, \cite{Hoare:2012nx} is a recent example. It is part of the use of integrability techniques to understand the AdS/CFT correspondence \cite{Integrability}. Finally, the possibility of solving the equations of motion using algebro-geometric methods is also known \cite{BB,BBook,ThF,FK} in the mathematical literature related to minimal areas surfaces and the theory of solitons.

\section{A simple example}\label{sec2}

\begin{figure}
  \centering
  \includegraphics[width=10cm]{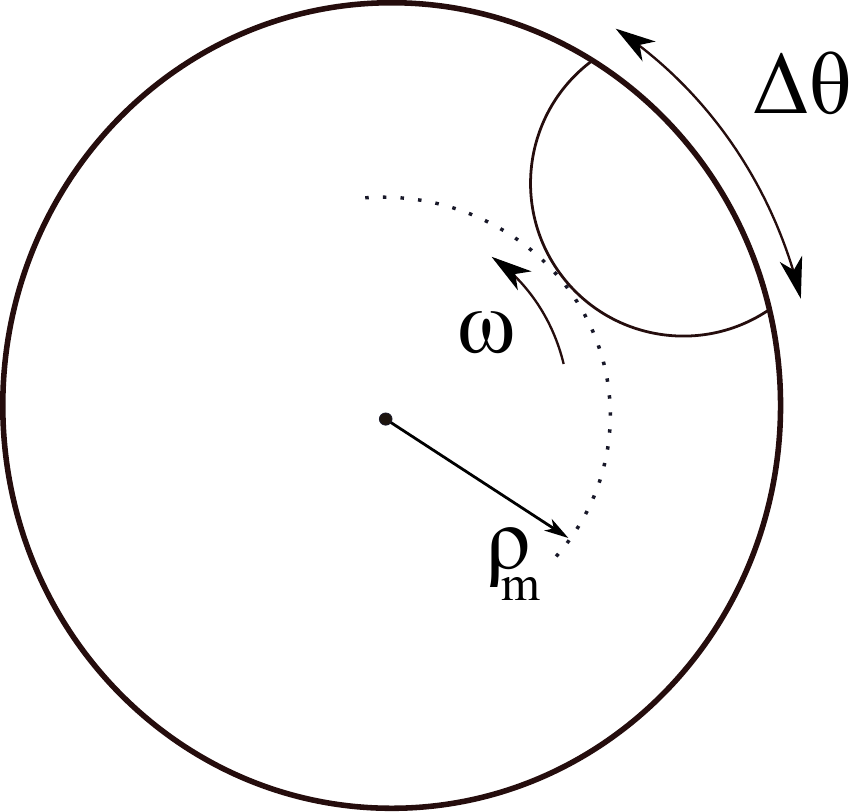}
  \caption{String ending on the boundary and rotating rigidly. The parameters that determine the solution are the angular velocity $\omega$ and the angular separation between the end points $\Delta \theta$. Alternatively one can use the angular velocity $\omega$ and the position $\rho_m$ of the point closest to the center.}
  \label{rigid1}
  \end{figure}

In this section, the simple case of a rigidly rotating string from reference \cite{Irrgang:2009uj} is revisited; namely, the situation depicted in fig.\ref{rigid1}.
The conformal factor in the world-sheet metric is seen to obey a cosh-Gordon, Liouville or 
sinh-Gordon equation depending on the relation between the angular velocity $\omega$ and the separation $\Delta \theta$ between the end points. 
Defining $\hat{\omega}=1-\frac{\Delta \theta}{\pi}$, the equation is cosh-Gordon if $\omega<\hat{\omega}$, Liouville if $\omega=\hat{\omega}$, or sinh-Gordon if $\omega>\hat{\omega}$.
Interestingly, the Liouville case where $\omega=\hat{\omega}=1-\frac{\Delta \theta}{\pi} $ is particularly simple and the shape of the string can be written in terms of trigonometric functions.

 Let us proceed now to describe the solution. The manifold \ads{3} can be defined as a subspace of $R^{2,2}$ given by the constraint
 \beq
X_\mu X^\mu =  X_1^2 + X_2^2 -X_3^2 - X_4^2 =1 ,
 \label{Ads}
 \eeq
 written in terms of the coordinates $X^\mu$ ($\mu = 1,2,3,4$). The action for a Lorentzian world-sheet in conformal gauge is given by
 \begin{equation}
 	S = -\frac{1}{2} \, T \int d\tau \, d\sigma \, (\pds{X}^\mu \pds{X}_\mu - \pdt{X}^\mu \pdt{X}_\mu  + \Lambda(X^\mu X_\mu + 1))
 \end{equation}
 where the Lagrange multiplier $\Lambda$ enforces the embedding constraint.  The equations of motion are
 \begin{equation}
 	\partial_\sigma^2{X}^\mu - \partial_\tau^2{X}^\mu = \Lambda X^\mu,
 	\label{eom1}
 \end{equation}
 and, due to the gauge choice, the solution must additionally satisfy the conformal constraints,
 \begin{eqnarray}
 	0 = \partial_\tau{X}^\mu\partial_\sigma{X}_\mu   &\qquad \qquad&   0 = \pdt{X}^\mu \pdt{X}_\mu + \pds{X}^\mu\pds{X}_\mu.
 \end{eqnarray}
 The same equations can be written using complex coordinates 
 \beq
 X = X_1+ i X_2,\ \ \  Y=X_3+i X_4, 
 \label{Ads2}
 \eeq
 or global coordinates defined through
 \beq
  X = \cosh\!\rho\, e^{it}, \ \ Y = \sinh\!\rho\, e^{i\theta}\, ,
 \label{Ads3}
 \eeq
 in which case, the \ads{3} metric takes the well-known form:
 \beq
 ds^2 = -\cosh^2\!\rho \ dt^2 + d\rho^2  + \sinh^2\! \rho\ d\theta^2 .
\label{ds}
 \eeq
 In the coordinates $(t,\rho,\theta)$ the rotating string depicted in fig.\ref{rigid1} is found by using the following ansatz:
 \beq
 \rho = \rho(\xi), \ t = \mu_0(\xi) + \omega_0 \tau, \ \theta=\mu_1(\xi) + \omega_1 \tau .
 \label{a1}
 \eeq
Here, $\xi=\alpha \sigma+\beta\tau$ and $(\sigma,\tau)$ are conformal coordinates on the world-sheet. Notice that these are open strings and,
therefore, there is no periodicity condition\footnote{For closed strings this ansatz can be problematic because $t$ depends explicitly on $\sigma$ 
through the function $\mu_0(\xi)$, which is generically not periodic as follows from the equations of motion.}. 
The ansatz is justified by writing the equations of motion, which are all solved if
\beq
 \mu'_0 = \frac{C_0}{\cosh^2\rho} + \frac{\beta\omega_0}{\alpha^2-\beta^2}, \ \ \ \ 
 \mu'_1 = \frac{C_1}{\sinh^2\rho} + \frac{\beta\omega_1}{\alpha^2-\beta^2},
\label{a2}
\eeq
and $\rho(\xi)$ obeys a differential equation better written in terms of $u=\sinh^2\rho$; namely,
\beq
u'{}^2 = \frac{4\alpha^2(\omega_0^2-\omega_1^2)}{(\alpha^2-\beta^2)^2}\ (u-u_1)(u-u_2)(u-u_3) .
\label{a3}
\eeq
Above, $C_{0,1}$ are constants such that $\omega_0 C_0=\omega_1 C_1$ and 
\beq
u_1 = -\frac{\omega_0^2}{\omega_0^2-\omega_1^2}, \ 
\ u_{2,3}=\half\left(-1\pm\sqrt{1+\frac{4(\alpha^2-\beta^2)^2C_0^2}{\alpha^2\omega_1^2}}\right)\, .
\label{a4}
\eeq
The solution is actually determined by only two constants: the angular velocity $\omega=\frac{\omega_1}{\omega_0}$
and the minimal value of $\rho$ denoted as $\rho_{m}\le\rho$ and given by $u_3=\sinh^2\rho_{m}$.

The world-sheet metric is
 \beq
 ds^2 = e^{2\phi} (d\tau^2-d\sigma^2) = e^{2\phi} d\tau_+d\tau_- \, ,
 \label{a5}
 \eeq
 with $\tau_\pm=\tau\pm\sigma$ and 
 \beq
 e^{2\phi}= \frac{\alpha^2}{\alpha^2-\beta^2}(\omega_0^2 + u (\omega_0^2-\omega_1^2)) .
 \label{a6}
 \eeq
The equations of motion then imply an equation for $\phi$:
 \beq
( \partial_\tau^2-\partial_\sigma^2) \phi = \frac{1}{\omega_0^2-\omega_1^2} \, e^{2\phi} 
    + \frac{\alpha^4(\omega_0^2-\omega_1^2)}{(\alpha^2-\beta^2)^2} \frac{(u_1-u_2)(u_3-u_1)}{e^{2\phi}}\ .
 \label{a7}
 \eeq
Up to a simple rescaling, this is the cosh-Gordon if $u_1>u_2$, Liouville if $u_1=u_2$ and sinh-Gordon equation if $u_1<u_2$. 
The reason being that all other factors are positive since $u_3>u_1$ and $\omega=\frac{\omega_1}{\omega_0}<1$.
 
\begin{figure}
  \centering
  \includegraphics[width=10cm]{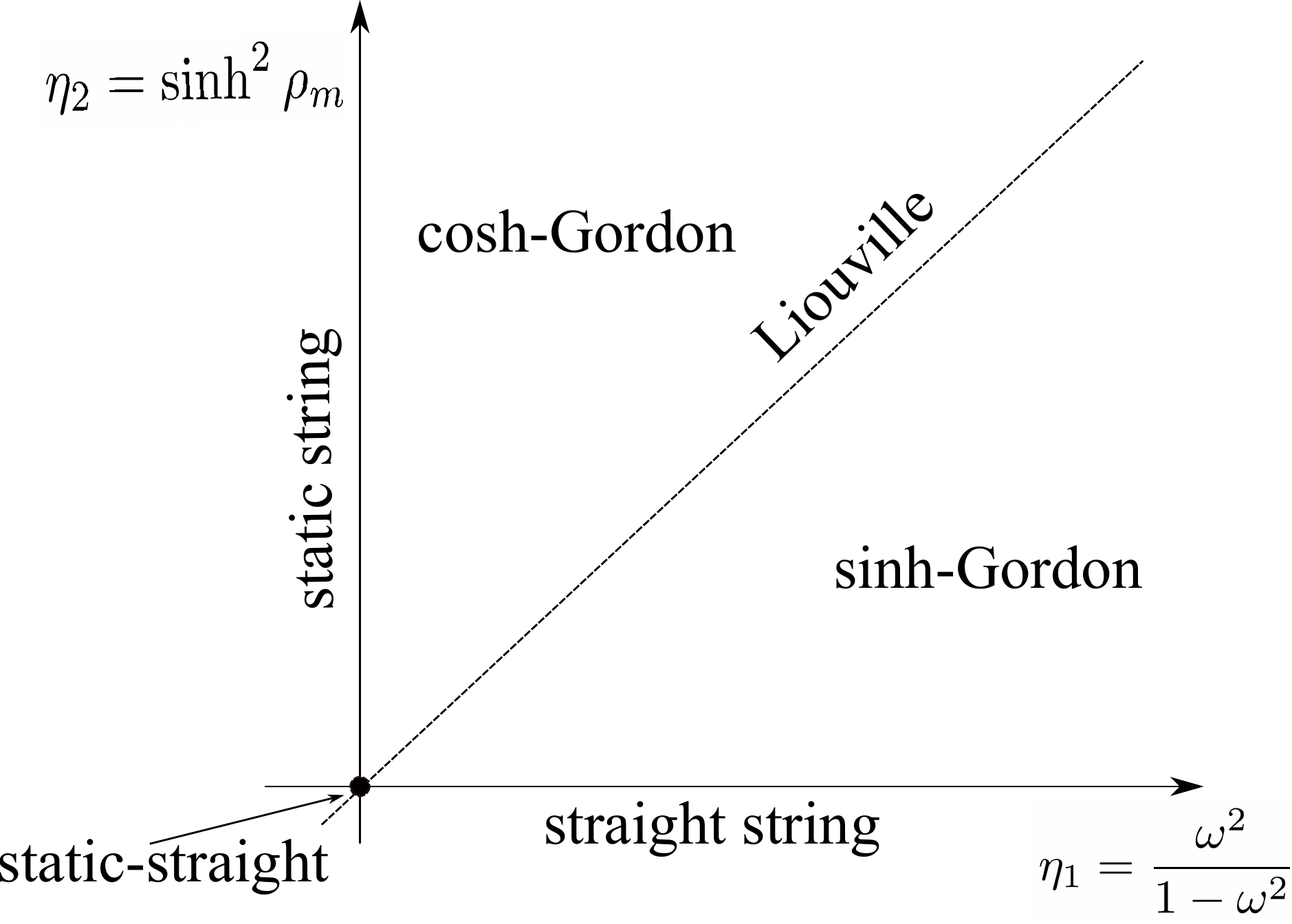}
  \caption{Depending on the relation between the angular velocity $\omega$ and the position of the lowest point of the string $\rho_{m}$, the equation obeyed by the world-sheet conformal factor is cosh-Gordon, Liouville or sinh-Gordon. The horizontal axis corresponds to a straight string ($\rho_m=0$) and the vertical one to a static string ($\omega=0$).}
  \label{rigid2}
  \end{figure}

A more physical picture is obtained by computing the angular separation between the end points of the string (at fixed time $t$). Introducing the constants $\eta_{1,2}=-1-u_{1,2}$, we find
 \beq
 \Delta\theta = \int_{1}^\infty \sqrt{\frac{(1+\eta_2)(1+\eta_2 u +\eta_1)}{(1+\eta_1)(1+\eta_2u+\eta_2)}} \frac{du}{u(1+u\eta_2)\sqrt{u-1}} .
 \label{a8}
 \eeq
A simple analysis of the integrand reveals that $\Delta \theta$ decreases with increasing $\eta_1$ and $\eta_2$.
When $\eta_1=\eta_2$, the separation simplifies to
 \beq
 \Delta \theta = \pi (1-\omega)\, .
 \label{a9}
 \eeq
Therefore, defining
 \beq
 \hat{\omega} = 1-\frac{\Delta\theta}{\pi},
 \label{a10}
 \eeq
the equation for $\phi$ reduces to Liouville if $\omega=\hat{\omega}$ .
After noting that $\eta_2=u_3=\sinh^2\rho_{m}$ and $\eta_1 = \frac{\omega^2}{1-\omega^2}$, the situation can be summarized by fig.\ref{rigid2}.
 
 The particular case leading to the Liouville equation (\ie\ $u_1=u_2$) is quite simple since two roots coincide and the shape of the string can be determined in terms of trigonometric functions:
 \beq
 \theta = \arccos\left(\frac{\sinh \rho_{m}}{\sinh\rho}\right) - \tanh\rho_{m}  \arccos\left(\frac{\cosh \rho_{m}}{\cosh\rho}\right)\, .
 \label{a11}
 \eeq
 It is also convenient to write the solution in the coordinates $(X,Y)$ defined in eq.(\ref{Ads3}):
 \beqa
 X &=& \frac{1}{\sqrt{\cos2\gamma}}\left(\frac{1+i\sigma_-}{\sqrt{1+\sigma_-^2}}\right)^{2\cos\gamma} \left(\cos\gamma + \frac{i(\sigma_--\sigma_+)}{1+\sigma_+\sigma_-}\right) , \label{a13}\\
 Y &=& \frac{1}{\sqrt{\cos2\gamma}}\left(\frac{1+i\sigma_-}{\sqrt{1+\sigma_-^2}}\right)^{2\sin\gamma} \left(\sin\gamma + \frac{i(\sigma_--\sigma_+)}{1+\sigma_+\sigma_-}\right) . \label{a14}
 \eeqa
 Here we used world-sheet coordinates:
 \beq
 \sigma_\pm = \tan\left(\half A(\alpha\pm\beta)(\sigma\pm\tau)\right), \ \ \ \ A=\frac{\alpha}{\alpha^2-\beta^2}\sqrt{\omega_0^2+\omega_1^2} .
 \label{a15}
\eeq
Finally, in the original coordinates $(\sigma,\tau)$, the conformal factor is
\beq
e^{2\phi} = \frac{A(\alpha^2-\beta^2)}{\cos^2A\xi}.
\label{a16}
\eeq
Recalling that $\xi=\alpha\sigma+\beta\tau$, it is easy to check that $\phi$ obeys the Liouville equation: 
\beq
(\partial_\sigma^2-\partial_\tau^2) \phi = A e^{2\phi} .
\label{a17}
\eeq
Here, $A$ is the one defined in eq.(\ref{a15}). 

As discussed later in this paper, the rigidly rotating string for $\omega=\hat{\omega}$ is a particular case of the solutions found by Mikhailov in \cite{Mikhailov:2003er}.
  
\section{The general case}\label{sec3}

In the previous section, a rigidly rotating string demonstrated that the conformal factor can obey the cosh-Gordon, Liouville, or sinh-Gordon equations. 
Now, we consider the general, non-rigid, case. It is convenient to use embedding coordinates,
\beq
X_1^2 + X_2^2 -X_3^2 - X_4^2 =1 ,
\label{a18} \eeq
and arrange them in a $2\times 2$ real matrix
\beq
\X = \left(\begin{array}{cc}X_1+X_3&X_4+X_2\\ X_4-X_2&X_1-X_3\end{array}\right)\, .
\label{a19} \eeq
This matrix obeys $\det\X=1$, \ie\ $\X\in SL(2,\mathbb{R})$.
Consider the decomposition
\begin{equation}
\mathbb{X} = A_1 A_2^{-1}\, ,
\end{equation}
with $\det A_a=1$, $a=1,2$. There is now a redundancy in the description and, as a result, a gauge symmetry
\beq
A_a \rightarrow A_a\, \cU(\tau_+,\tau_-)\, ,
\label{a20} \eeq
leaves $\X$ invariant. Define now two one-forms:
\begin{equation}
J_a = A_a^{-1} d A_a,  \ \ \ a=1,2\, .
\end{equation}
It follows, with no summation on $a$ implied, that
\begin{equation}
\tr J_a =0, \ \ \ dJ_a +J_a\wedge J_a =0 .
\end{equation}
To be precise, the conventions used for differential forms in coordinates $\tau_\pm=\tau\pm\sigma$ are:
\begin{eqnarray}
a &=& a_+ d\tau_+ + a_- d\tau_-, \\ 
da &=& (\partial_- a_+ - \partial_+ a_-)\, d\tau_-\wedge d\tau_+, \\
a\wedge b &=& (a_- b_+ - a_+ b_- ) d\tau_-\wedge d\tau_+, 
\end{eqnarray}
\begin{equation}
  (*a)_+ = a_+, \ (*a)_- = -a_-, \ \ *a\wedge b = -a\wedge *b, \ \ \ **a=a.
\end{equation}
The equations of motion (\ref{eom1}) become
\begin{equation}
*d\!*\!d\mathbb{X} = 2\Lambda\, \mathbb{X} .
\end{equation}
Expanding the derivatives in terms of the currents yields:
\begin{equation}
 J_1 \wedge *J_1 + *J_1 \wedge J_2 - J_1\wedge *J_2 - *J_2\wedge J_2 + d*J_1 - d*J_2 = 2\Lambda .
\end{equation}
 Since the currents are two by two traceless matrices the traceless part can be extracted by using commutators; namely,
\begin{equation}
 A\wedge B \xrightarrow{\mbox{traceless part}} \half (A\wedge B +B\wedge A) .
\end{equation}
Here, the right-hand side is the traceless part of the left-hand side. Thus, we get a simple equation
\begin{equation}
d*(J_1-J_2) + *J_1\wedge J_2 + J_2\wedge *J_1 = 0 .
\end{equation}
Also, the equations of motion can be written in terms of the currents:
\begin{eqnarray}
 dJ_1 + J_1\wedge J_1 &=& 0 ,\\
dJ_2 + J_2\wedge J_2 &=& 0 ,\\
d*(J_1-J_2) + *J_1\wedge J_2 + J_2\wedge *J_1 &=& 0 .
\end{eqnarray}
These equations need to be supplemented with the constraints $\det(\partial_+ \mathbb{X}) = 0 = \det(\partial_- \mathbb{X})$; 
equivalently, 
\begin{equation}
\det (J_{1+} - J_{2+}) =0 = \det (J_{1-} - J_{2-}) .
\end{equation}
 At this point, it is convenient to define two new currents,
 \beq
 \cA = \half (J_1-J_2) , \ \ \ \cB = \half(J_1+J_2)\ ,
 \label{b1} \eeq
in terms of which the equations of motion can be rewritten:
\beqa
 d\cA + \cA \wedge \cB + \cB\wedge \cA &=& 0              ,\\
 d(*\cA) + (*\cA) \wedge \cB + \cB\wedge (*\cA) &=& 0     ,\\
  d\cB + \cB \wedge \cB + \cA \wedge \cA &=& 0            ,\\
  \det(\cA_+) = \det(\cA_-) &=&0                          ,\\
  \tr \cA = \tr \cB &=&0                                  .
\label{b2} \eeqa
 A flat current can be found as a linear combination:
\begin{eqnarray}
 a &=& \alpha \cA + \beta *\!\cA + \gamma \cB,\\
 da+a\wedge a &=& 0 ,
\end{eqnarray}
which in addition satisfies
\begin{equation}
\tr(a) =0 .
\end{equation}
There is a one parameter family of non-trivial solutions given by $(\alpha+\beta)(\alpha-\beta)=1$, $\gamma=1$.
It can be conveniently parameterized in terms of the spectral parameter $\lambda$ as $\alpha+\beta=\lambda$, $\alpha-\beta=\frac{1}{\lambda}$. Thus:
\beqa
a_+ &=& \lambda \cA_+ + \cB_+             ,\\
a_- &=& \frac{1}{\lambda} \cA_- + \cB_-  .
\label{b3} \eeqa
Since $\cA$ and $\cB$ are real but $\lambda$ is generically complex, the flat current $a$ satisfies the reality condition 
\beq
\overline{a(\lambda)} = a(\bar{\lambda}) .
\label{b4} \eeq
It is also useful to note that $J_1= a(1)$, $J_2=a(-1)$. 
Returning to the current $\cA$, and expanding it in terms of the Pauli matrices $\sigma_{a=1,2,3}$:
\beqa
\cA_+ &=& n_1 \sigma_1 +n_2 i\sigma_2 + n_3 \sigma_3  ,\\
\cA_- &=& \tilde{n}_1 \sigma_1 +\tilde{n}_2 i\sigma_2 + \tilde{n}_3 \sigma_3 ,
\label{b5} \eeqa
the condition $\det \cA_\pm=0$ implies that $n$, $\tilde{n}$ are light-like vectors, \ie\
\beq
 n_2^2 - n_1^2 -n_3^2 =0 ,
\label{b6} \eeq
and the same for $\tilde{n}$. The gauge symmetry $A_a \rightarrow A_a\, \cU(\tau_+,\tau_-)$, in terms of $\cA$, gives 
$\cA_a \rightarrow \cU(\tau_+,\tau_-)^{-1} \cA_a\, \cU(\tau_+,\tau_-)$ which is an 
$SL(2,\mathbb{R})=SO(2,1)$ rotation of the vectors $n$, $\tilde{n}$. Assuming that $n\neq\tilde{n}$, they can always be put in the form:
\beq
 n = e^{\alpha} (1,1,0), \ \ \tilde{n} = e^{\alpha} (-1,1,0) ,
\label{b7} \eeq
where $\alpha(\tau_+,\tau_-)$ is a real function. In this way, the current is
\beq
\cA = e^{\alpha} (d\tau_+ \sigma_+ + d\tau_- \sigma_-) ,
\label{b8} \eeq
where $\sigma_+ = \left(\begin{array}{cc}0&1\\0&0\end{array}\right)$, $\sigma_- = \left(\begin{array}{cc}0&0\\1&0\end{array}\right)$. 
The equation for $\cB$ determines that
\beqa
\cB_+ &=& \half \partial_+\, \alpha\, \sigma_3  + u(\tau_+) e^{-\alpha}\, \sigma_-  ,\\
\cB_- &=& -\half \partial_-\, \alpha\, \sigma_3 +  v(\tau_-) e^{-\alpha}\, \sigma_- ,
\label{b9} \eeqa
where $u(\tau_+)$ and $v(\tau_-)$ are arbitrary functions and, in addition, $\alpha$ satisfies
\beq
\half \partial_{+-} \alpha  = e^{2\alpha} - e^{-2\alpha} u(\tau_+) v(\tau_-) .
\label{b10} \eeq
 By an appropriate change of coordinates $\tau_\pm$ and a redefinition of $\alpha$, one can set $u$ and $v$ to a constant except at those points
 where they vanish. In the rest of the paper we consider only the case where $uv$ is constant everywhere and can therefore be set equal to either, $-1$, $0$ or $+1$.  
 The case where 
 $uv$ vanishes at a finite set of points will not be considered since we do not know how to write the general solution in those cases. 
 
 The flat connection can be written as
 \beq
 a_+ = \left(\begin{array}{cc}\half\partial_+\alpha & \lambda e^{\alpha}\\ ue^{-\alpha} & -\half \partial_+\alpha\end{array}\right), \ \ \  
 a_- = \left(\begin{array}{cc}-\half\partial_-\alpha & v e^{-\alpha}\\ \frac{1}{\lambda}e^{\alpha} & \half \partial_-\alpha\end{array}\right),
\label{b11} \eeq
which is valid in general even if only the case of constant $u,v$ is considered here. Since $a$ is flat, we can solve the linear problem
\beq
d \Psi(\lambda;\tau_+,\tau_-) = \Psi(\lambda;\tau_+,\tau_-) a .
\label{b12} \eeq
Moreover, since $J_1= a(1)$, $J_2=a(-1)$, we have $A_1=\Psi(1)$, $A_2=\Psi(-1)$; namely,
\beq
\X = \Psi(1) \Psi(-1)^{-1} .
\label{b13} \eeq
Therefore, the strategy is to solve the equation for $\alpha$, replace it in the flat current, solve the linear problem, and reconstruct the solution $\X$. 
In fact, a family of solutions satisfying the equations of motion and the constraints can be introduced:
\beq
\X (\lambda) = \Psi(\lambda) \Psi(-\lambda)^{-1} .
\label{b14} \eeq
Furthermore, the reality condition is satisfied if $\lambda\in\mathbb{R}$.

\section{Liouville case ($uv=0$)}\label{sec4}

When $uv=0$, equation (\ref{b10}) reduces to the Liouville equation:
\beq
\half \partial_{+-} \alpha =  e^{2\alpha} .
\label{a23} \eeq
The general solution to this equation is
\beq
\alpha = \half \ln\left(\frac{2\partial_+f(\tau_+) \partial_-\tilde{f}(\tau_-)}{(1-f(\tau_+)\tilde{f}(\tau_-))^2}\right) .
\label{a24} \eeq
In this case, to reconstruct the solution $X_\mu(\tau_+,\tau_-)$, we can side-step the procedure described in the previous section and directly solve
the linear problem
\beq
\partial_{+-} X^\mu = e^{2\alpha} X^\mu .
\label{a25} \eeq
Using separation of variables (and taking as coordinates $f,\tilde{f}$) or by any other technique, the general solution is found to be
\beq
X^\mu = \left[\frac{1+f\tilde{f}}{1-f\tilde{f}} h^\mu_+(\tau_+) + \frac{f}{\partial_+f} \partial_+ h_+^\mu(\tau_+)\right] +
        \left[\frac{1+f\tilde{f}}{1-f\tilde{f}} h^\mu_-(\tau_-) + \frac{\tilde{f}}{\partial_-\tilde{f}} \partial_- h_-^\mu(\tau_-)\right] .
\label{a26} \eeq
where $h^\mu_+(\tau_+)$ and $h^\mu_-(\tau_-)$ are arbitrary functions.
Although this satisfies the equations of motion, it also needs to satisfy the constraints 
\beq
(\partial_+ X^\mu)^2=0=(\partial_- X^\mu)^2, \ \ \ X^2=1 .
\label{a27} \eeq
Although particular solutions exist, it appears that generic solutions cannot be found in the case where both $h_+$ and $h_-$ are different from zero. In this paper,
only the case where $h^\mu_-=0$ is considered and the solution is
\beq
X^\mu = \left[\frac{1+f\tilde{f}}{1-f\tilde{f}} h^\mu_+(\tau_+) + \frac{f}{\partial_+f} \partial_+ h_+^\mu(\tau_+)\right] .
\label{a28} \eeq
The constraints then reduce to
\beqa
 h^2 &=& 0 ,\\ 
 (\partial_+ h)^2 &=& (\partial_+ \ln f)^2 ,\\
 (\partial_{++} h)^2 &=& (\partial_{++}\ln f)^2  - (\partial_+ \ln f)^4 .
 \label{a29} \eeqa
 These are relatively easy to study by noting that different choices of the functions $f(\tau_+)$, $\tilde{f}(\tau_-)$ are related by reparameterizations of the world-sheet and, therefore, equivalent. 
In this way, we can choose for example
 \beq
 f = e^{i\tau_+}, \ \ \tilde{f} = -e^{i\tau_-} ,
 \label{a30} \eeq
without loosing generality. Now, the solution can be written as
\beq
X^\mu = \tan\sigma\  h_+^\mu(\tau_+) + \partial_+ h_+^\mu ,
\label{a31} \eeq
with 
\beq
 h_+^2 = 0, \ \  (\partial_+h_+)^2 = 1 , \ \ \ (\partial_{++}h_+)^2 = 1.
 \label{a32} \eeq
 The string ends at the boundary on two curves determined by $\sigma=\pm \frac{\pi}{2}$. The curves are given by 
 \beq
  X^{\mu} = \pm h^\mu_+(\tau\pm\frac{\pi}{2}) .
\label{a33} \eeq
Notice that $X^2=0$ which is a parameterization of the boundary of \ads{} space. Furthermore, these solutions are valid for any \ads{} and not only \ads{3}. In fact they are already known since they were found by Mikhailov in \cite{Mikhailov:2003er}. 
In particular, they contain the solution corresponding to the rotating string described in the first section. To show that, start by parameterizing the boundary curves as
\beq
 h_+^\mu = \left(\begin{array}{c}\cos t \\ \sin t\\ \sin\theta \\ \cos \theta\end{array}\right) .
\label{a34} \eeq
 Next, observe that, according to eq.(\ref{a33}), the two boundary curves are the same up to a shift by $\pi$ in the parameter $\tau$ and a change in sign of $h_+^\mu$, which is equivalent to a
 shift $t\rightarrow t+\pi,$ , $\theta\rightarrow \pi + \theta$. Since the shift in parameter $\tau$ does not change the shape of the curve, one curve is obtained from the other by rotating 
 $\theta$ by $\pi$ and shifting $t$ also by $\pi$. Therefore, the shape of one boundary curve can be chosen arbitrarily, which determines the functions $h^\mu_+(\tau_+)$ and also the shape of the other boundary curve. 

\begin{figure}
  \centering
  \includegraphics[width=10cm]{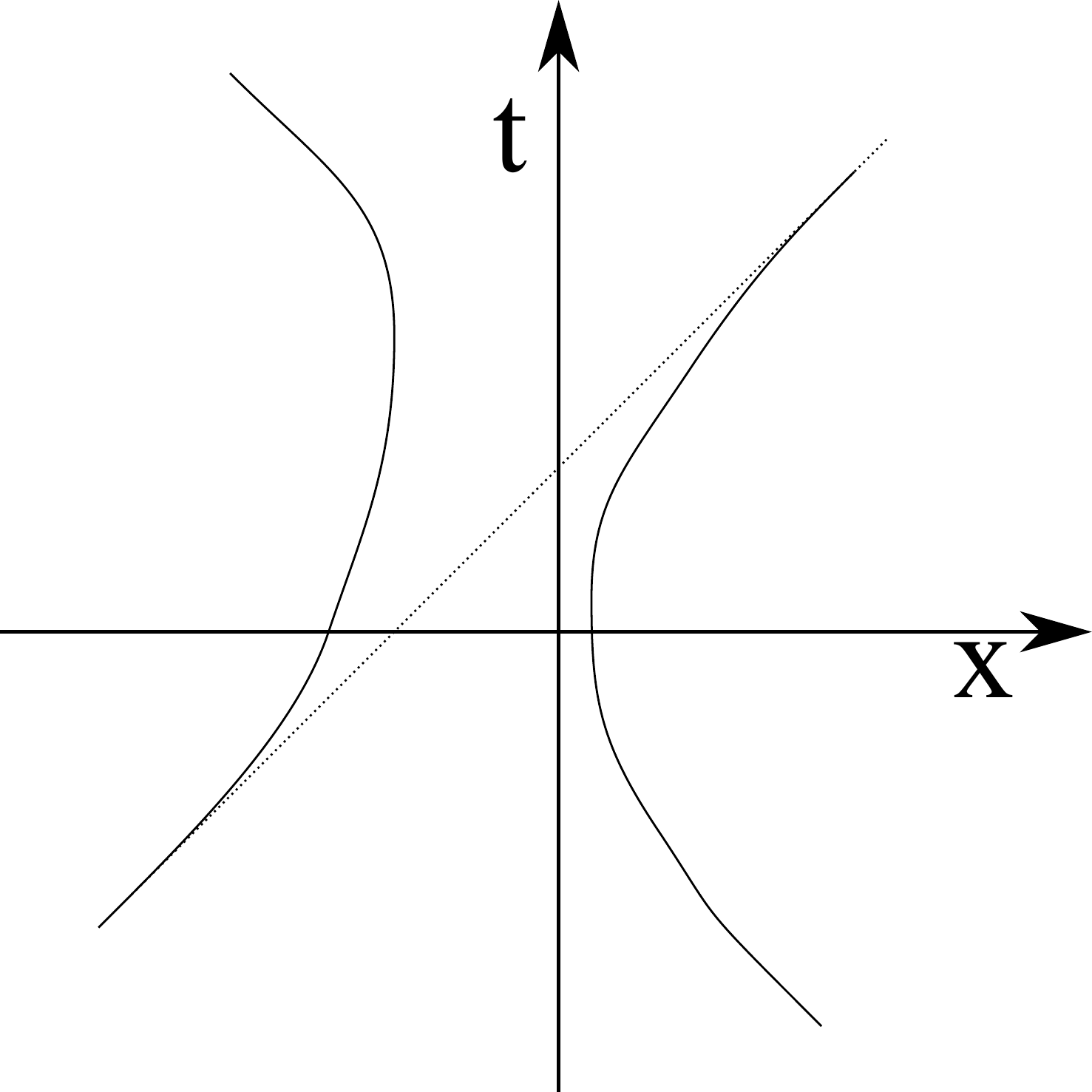}
  \caption{In Poincare coordinates, the Liouville case generically corresponds to a string ending on two arbitrary paths which are asymptotically light-like. The only condition is that the
  outgoing direction of one of them is equal to the incoming direction of the other one.}
  \label{Liouville}
  \end{figure}

For the rigidly rotating string, we have that one curve is determined by
\beq
 \theta_1 = \omega t .
 \label{a35} 
 \eeq
 Consequently, the other is
 \beq
 \theta_2 + \pi = \omega( t+\pi) ,
 \label{a36} 
 \eeq
 This implies that
 \beq
 \Delta \theta = \theta_1 - \theta_2 = \pi (1-\omega) ,
 \label{a37} 
 \eeq
 which is precisely the result in eq.(\ref{a9}). Physically, this means that a ray of light emitted from the quark reaches the anti-quark exactly after a time $t=\pi$. If it reaches earlier or
 later we then have the cosh or sinh Gordon equation.  In Poincare coordinates, the two end points of the string describe arbitrary trajectories that are asymptotically null. In this case, it should be noted that both trajectories are arbitrary since they come from different portions of the trajectory in global \ads. The only condition is that one asymptotes to a null direction that
is equal to the incoming direction of the other one. See fig.\ref{Liouville}.

\section{Sinh/cosh Gordon case ($uv=\pm1$)}\label{sec5}

 The other case to consider is when the world-sheet conformal factor obeys the sinh or cosh-Gordon equation, \ie\ eq.(\ref{b10}) with $uv=\pm 1$. The solution can be written in terms of theta functions
 associated with a hyperelliptic Riemann surface. The reality condition implies that the Riemann surface can be thought as two planes connected by a set of cuts symmetric under the interchange
 $\lambda\leftrightarrow \bar{\lambda}$. In the examples considered here, the cuts are on the real axis although other situations are possible.
 
To be concrete, the hyperelliptic Riemann surface of genus $g$ is defined by the equation
 \beq
 \mu^2 = \prod_{i=1}^{2g+1} (\lambda-\lambda_i) ,
 \label{a38} \eeq
 where $(\mu,\lambda)$ parameterize $\mathbb{C}^2$. A basis of 1-cycles $(a_i,b_i)$ is chosen together with a basis of normalized holomorphic differentials $\omega$. 
  Points on the Riemann surface are denoted as $p_i$ whereas their projection on the complex plane is denoted as $\lp{i}$. Obviously, each $\lp{i}$ corresponds to two points on the Riemann surface (upper and lower sheets) except for the branch points $\lambda_i$. 
  Two of the branch points are going to be singled out and denoted as $p_1$ and $p_3$. Take a path from $p_1$ to $p_3$ on the upper sheet and close it by tracing the same path backwards on the lower sheet. 
  When written in the basis $(a_i,b_i)$, the closed path $\mathcal{C}_{13}$ defines two integer vectors $\Delta_1$ and $\Delta_2$ such that
  \beq
   \mathcal{C}_{13} = \Delta_{2i} a_i + \Delta_{1i} b_i \, .
  \label{a39} \eeq  
  These vectors, together with the periodicity matrix of the Riemann surface, define a theta function with characteristics:
   \beq
   \hat{\theta}(\zeta) = \theta\left[\begin{array}{c}\Delta_1\\ \Delta_2\end{array}\right](\zeta)  , \ \ \ \ \zeta\in \mathbb{C}^g
   \label{a40} \eeq
This function and the usual theta function $\theta(\zeta)$ without characteristics determine a function $\alpha(\tau_+,\tau_-)$ 
    \beq
    e^{\alpha} = C_\alpha \frac{\theta(\zeta)}{\hat{\theta}(\zeta)} , \ \ \ \ \   \zeta = C_-\, \hat{\omega}_1\, \tau_- + C_+ \,\hat{\omega}_3\, \tau_+
    \label{a41} \eeq
    which solves the sinh/cosh-Gordon equation.
    The constant $C_\alpha$ is such that $C_\alpha^4=1$ and should be chosen such that the right hand side is real and positive. This is always possible since the reality conditions ensure that $\theta(\zeta)$ is real and $\hat{\theta}(\zeta)$ is either real or purely imaginary.  Moreover, as discussed later, this is the only reality condition needed. As explained below, once $\alpha$ is real, a real solution for $X^\mu(\sigma,\tau)$ can always be obtained.
    More details on the notation and derivations can be found in the appendix including the definition of the vectors $\hat{\omega}_{1,3}$.
     Finally, the constants $C_\pm$ are such that $C_\pm^2=1$ and should be chosen such that
    \beq
      C_+ C_- = e^{i\pi\Delta_2^t \Delta_1}\, C_\alpha^2 .
    \label{a47} \eeq
    With these definitions, it is just a matter of algebra to check that $\alpha$ satisfies the equation
   \beq
   \partial_{+-}\alpha = e^{2\alpha} - e^{i\pi\Delta_2^t \Delta_1} e^{-2\alpha} ,
   \label{a48} \eeq
   as shown in the appendix.
    Consequently, we have to identify
    \beq
    uv = e^{i\pi \Delta_1^t \Delta_2} .
    \label{a49} \eeq
    In what follows, it is convenient to choose
    \beq
     u=1, v =  e^{i\pi \Delta_1^t \Delta_2}=\pm 1
     \label{a50} \eeq
    Summarizing, if we choose $\lp{1}$, $\lp{3}$ such that $\Delta_2^t\Delta_1$ is even, we have the sinh-Gordon equation. If we choose them such that $\Delta_2^t\Delta_1$ is odd we obtain the cosh-Gordon one. 
   
    It should be noted that there is a different solution 
        \beq
        e^{\alpha} = C_\alpha \frac{\theta(i\zeta)}{\hat{\theta}(i\zeta)} ,
        \label{a51} \eeq
   which is interesting but simply related by $\zeta\rightarrow i \zeta$ to the previous one. For that reason it is not considered further. 
    
   The next step is to write the flat current $a(\lambda)$ and find the matrix $\Psi$ that satisfies
   \beq
   d\Psi (\lambda) = a(\lambda) \Psi(\lambda) .
   \label{a52} \eeq
    The matrix $\Psi$ is a two by two matrix that we write as
   \beq
    \Psi =\left(\begin{array}{cc}
    \psi_1 & \psi_2 \\
    \tilde{\psi}_1 & \tilde{\psi}_2
    \end{array}\right) ,
   \label{a52b} \eeq
    where $(\psi_1,\psi_2)$ and $(\tilde{\psi}_1, \tilde{\psi}_2)$ are two linearly independent solutions of the equation
     \beq
        d\psi (\lambda) = a(\lambda) \psi(\lambda) ,
     \label{a53} \eeq
     where $\psi$ is now a two dimensional row vector. Using expression (\ref{b11}) for $a(\lambda)$, the equations can be rewritten:
     \beqa
      \partial_+ \psi_1 &=& \half \partial_+ \alpha \psi_1 + u e^{-\alpha} \psi_2 ,\\
      \partial_+ \psi_2 &=& \lambda e^{\alpha} \psi_1 - \half \partial_+ \alpha \psi_2 ,\\
      \partial_- \psi_1 &=& -\half \partial_- \alpha \psi_1 + \frac{1}{\lambda} e^{\alpha} \psi_2 ,\\
      \partial_- \psi_2 &=& v e^{-\alpha} \psi_1 + \half \partial_- \alpha \psi_2 .
      \label{a54} \eeqa
     Here $u$ and $v$ are as in eq.(\ref{a50}). It is important to note that once $\alpha$ is real and the spectral parameter $\lambda$ is taken real, $\psi_{1,2}$ obey real equations and, therefore, can always be taken to be real. 
     For example, given a solution, its real and imaginary part are real and also solve the equations.
     
     Now we can proceed to solve the equations. One way to do it is to convert them into an equation for the ratio $\psi_1/\psi_2$ which can then be solved using the identities (\ref{a99}). Afterwards, one can solve for $\psi_{1,2}$ individually, again using eqns.(\ref{a99}). The result is
     \beqa
             \psi_1 &=& C_1 e^{\half\alpha} \frac{\theta(\zeta+\int_1^4)}{\theta(\zeta)} \ e^{\mu_+\sigma_++\mu_-\sigma_-} ,\\
             \psi_2 &=& C_2 e^{\half\alpha} \frac{\theta(a+\int_4^1)}{\hat{\theta}(a+\int_4^1)} \frac{\hat{\theta}(\zeta+\int_1^4)}{\theta(\zeta)} \ e^{\mu_+\sigma_++\mu_-\sigma_-} \label{a55} ,
     \eeqa
      where the constants $C_{1,2}$, $\mu_\pm$ and $\lambda$ need to be determined. By comparing with the theta function identities found in the 
      appendix, eq.(\ref{a99}), one finds that the spectral parameter is given by
      \beq
       \lambda(\lp{4}) = C_\lambda\, \frac{\theta^2(a+\int_4^1)}{\hat{\theta}^2(a+\int_4^1)} = A_0 \frac{\lp{4}-\lp{1}}{\lp{4}-\lp{3}}\ .
       \label{a56} \eeq
       That the ratio of theta functions simplifies as in the last equation is explained in the appendix. Here $A_0$ is a constant independent of $\lp{4}$ but dependent on the other parameters of the solution; namely, the Riemann surface and the points $\lp{1}$, $\lp{3}$. 
       The other constants are such that
       \beq
        C_\lambda = v, \ \ \ C_2 = C_\alpha C_+ C_1 .
       \label{a57} \eeq
       The constants $\mu_\pm$ turn out to be
       \beqa
           \mu_+ &=& -C_+ \frac{\hat{\theta}(a)}{D_3\theta(a)} D_3 \ln \frac{\hat{\theta}(a)}{\hat{\theta}(a+\int_4^1)} , \\
           \mu_- &=& -C_- \frac{\hat{\theta}(a)}{D_1\theta(a)} D_1 \ln \frac{\hat{\theta}(a)}{{\theta}(a+\int_4^1)} .
           \label{a58} \eeqa
       As discussed in the appendix, there are three possible cases:
       \begin{itemize}
       \item[1)] $v=+1$, $\theta,\hat{\theta}$ are real. The equation is sinh-Gordon and we can take $C_+=C_-=C_\lambda=1$ and $C_\alpha = \pm1$ such that $e^\alpha >0$. Finally $C_2 = C_\alpha C_1$.
       \item[2)]  $v=-1$, $\theta,\hat{\theta}$ are real. The equation is cosh-Gordon and we can take $C_+=1$, $C_-=C_\lambda=-1$ and $C_\alpha = \pm1$ such that $e^\alpha >0$. Finally $C_2 = C_\alpha C_1$.
        \item[3)]  $v=-1$, $\theta$ real,$\hat{\theta}$ purely imaginary. The equation is cosh-Gordon and we can take $C_+=C_-=1$, $C_\lambda=-1$ and $C_\alpha = \pm i$ such that $e^\alpha >0$. Finally $C_2 = C_\alpha C_1$.
        \end{itemize}
        As discussed above, $\psi_{1,2}$ obey real equations; therefore, if the previous solutions are complex, in fact, two solutions are obtained and are
        given by its real and imaginary part. If those are independent we are done. Otherwise, we have to find another linearly independent solution which can be easily done: 
       \beqa
         \tilde{\psi}_1 &=& A e^{\half\alpha} \frac{\theta(\zeta+\int_1^{\bar{4}})}{\theta(\zeta)} \ e^{\tilde{\mu}_+\sigma_++\tilde{\mu}_-\sigma_-} ,\\
         \tilde{\psi}_2 &=& A e^{\half\alpha} \frac{\theta(a+\int_{\bar{4}}^1)}{\hat{\theta}(a+\int_{\bar{4}}^1)} \frac{\hat{\theta}(\zeta+\int_1^{\bar{4}})}{\theta(\zeta)} \ e^{\tilde{\mu}_+\sigma_++\tilde{\mu}_-\sigma_-}  .
        \label{a59} \eeqa
   The notation $\bar{4}$ denotes the point $\bar{p}_4$ which is different from $p_4$ but has the same projection $\lp{4}$. That is, one is in the lower sheet and the other in the upper sheet.
   Notice that in eq.(\ref{a56}), the value of $\lambda$ depends only on $\lp{4}$ meaning that it does not matter if $p_4$ is in the upper or lower sheet; therefore, $(\psi_1,\psi_2)$ and $(\tilde{\psi}_1, \tilde{\psi}_2)$ satisfy the same differential equation. 
   Furthermore, define the constants
   $\tilde{\mu}_\pm$ by replacing $4\rightarrow \bar{4}$. 
   \begin{figure}
   \centering
   \includegraphics[width=12cm]{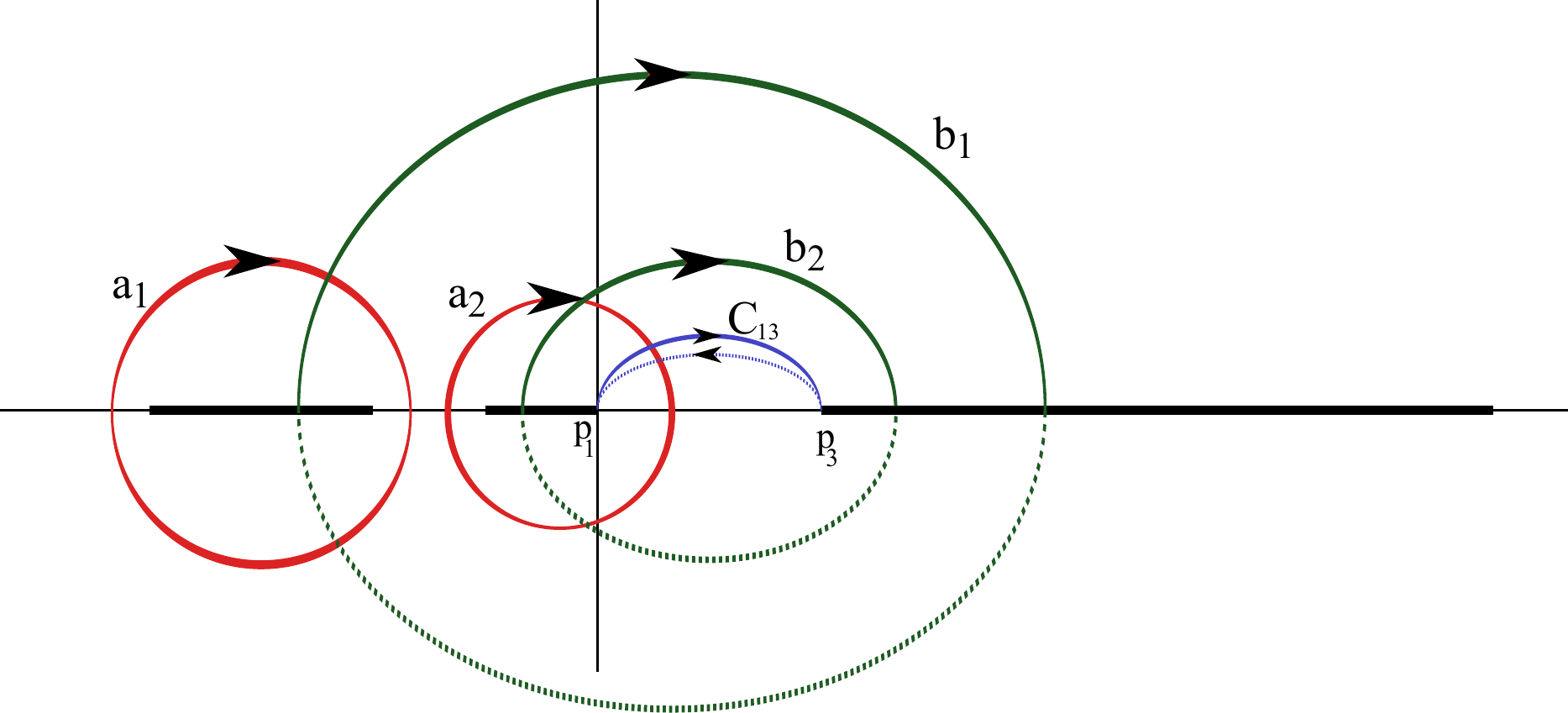}
   \caption{A simple example of Riemann surface with three cuts ($g=2$) and a choice of basis $(a_i,b_i)$ for the cycles. The cuts are on the real axis implying symmetry under $\lambda\leftrightarrow \bar{\lambda}$. Furthermore, under conjugation, $\bar{a}_i=-a_i$, $\bar{b}_i=b_i$. Two of the branch points are chosen as $p_1$ and $p_3$. They define the path $C_{13}$ which in this case is just $C_{13}=b_2$. Therefore, in this example, $\Delta_1=[0,1]$ and $\Delta_2=[0,0]$. }
   \label{cuts}
   \end{figure}
 In this way, we have found $\Psi(\lambda)$. However, we know that the solution is given by $\X=\Psi(\lambda)\Psi(-\lambda)^{-1}$ so we still need to find 
 $\Psi(-\lambda)$. Since $\Psi(-\lambda)$ satisfies the same equation, the solution is similar, we only need to find a point $\lp{4'}$ such that
 \beq
 \lambda' = C_\lambda\, \frac{\theta^2(a+\int_4^1)}{\hat{\theta}^2(a+\int_4^1)} = -\lambda .
 \label{a60} \eeq
 Given the formula
 \beq
 C_\lambda\, \frac{\theta^2(a+\int_4^1)}{\hat{\theta}^2(a+\int_4^1)} = A_0 \frac{\lp{4}-\lp{1}}{\lp{4}-\lp{3}} ,
 \label{a61} \eeq
 such point is simply given by
 \beq
 \lp{4'} = \frac{\lp{4}(\lp{1}+\lp{3})-2\lp{1}\lp{3}}{2\lp{4}-\lp{1}-\lp{3}} .
 \label{a62} \eeq
 Replacing $\lp{4}$ by $\lp{4'}$ everywhere, we find $\Psi(-\lambda)$ and with that the solution $\X(\tau_+,\tau_-)$. In fact, a family of solutions parameterized by a real parameter $\lambda$ (or equivalently the point $p_4$). Having taken care of choosing real solutions for $\Psi(\pm\lambda)$, the solution $\X$ is real. The constraints are also automatically satisfied by construction. The best way of understanding what type of solutions these represent is, perhaps, to work out an example as we do in the next section.

 \section{Examples}\label{sec6}
 
  To get an idea of the shape of the solutions, we give here two examples, one corresponds to the sinh-Gordon and the other to the cosh-Gordon case. 
  We give (approximate) numerical values for all the quantities appearing in the solution so that the interested reader can reproduce the results if he/she so desires.  
 For simplicity, the auxiliary Riemann surface is taken to be of genus $g=2$. The cuts are taken on the real axis as $[-2,-1]$, $[-\half,0]$ and $[1,\infty]$.
 The point $p_1$ is taken as $p_1=0$ and the point $p_3$ is $p_3=1$ for the sinh-Gordon case and $p_3=-1$ for the cosh-Gordon case. 
 The cycles $a_{1,2}$ and $b_{1,2}$ are taken as in fig.\ref{cuts} which determines (see the appendix 
 for details on the procedure)
\beqa 
 C_{ij} &= &\oint_{a_i} \frac{\lambda^{i-1}}{\mu(\lambda)}\ d\lambda  = \left(\begin{array}{cc} 3.691 & -5.042 \\ -5.042 & 1.351 \end{array}\right)_{ij} \ \ ,\\
 \tilde{C}_{ij} &= &\oint_{b_i} \frac{\lambda^{i-1}}{\mu(\lambda)}\ d\lambda  = \left(\begin{array}{cc} 1.351i & -5.042i \\ -3.691i & -1.351i \end{array}\right)_{ij} \ \ \ .
\label{a63} \eeqa
where
\beq
\mu^2(\lambda) = (\lambda+2) (\lambda+1) (\lambda+\half) \lambda (\lambda-1)
\eeq
 The periodicity matrix is given by
 \beq
  \Pi = \tilde{C}\, C^{-1} = \left(\begin{array}{cc} 1.155 i & 0.577 i \\ 0.577 i & 1.155 i \end{array} \right) .
 \label{a64} \eeq  
  The zero of the theta function $a$ is taken as the odd half period:
  \beq
  a=\half \vect{0}{1} + \half \Pi \vect{0}{1}=\vect{0.289 i}{\half + 0.577 i} .
  \label{a65} \eeq
  The normalized holomorphic differentials are
  \beq
  \omega_i = \frac{\lambda^{j-1}}{\mu(\lambda)}\, C^{-1}_{ji}\ \ \ \Rightarrow \ \ \ \oint_{a_j} \omega_i = \delta_{ij} .
  \label{a66} \eeq 
Now we consider two separate examples: one for sinh and the other for cosh-Gordon.
\subsection{Sinh-Gordon} 
For the sinh-Gordon case, we choose $p_1=0$, $p_3=1$ that determines the characteristic of $\hat{\theta}$ as:
 \beq
 \int_{p_1}^{p_3}\, \omega = \half \Delta_2 + \half \Pi \Delta_1  \ \ \ \Rightarrow \ \ \ \Delta_1 = \vect{0}{1},\ \ \ \Delta_2 =\vect{0}{0}
 \label{a67} \eeq
which is even and therefore leads to the sinh-Gordon equation. The constants are simply taken to be
\beq
C_\alpha = C_1 = C_2 = C_+ = C_- = 1 .
\label{a68} \eeq
The point $p_4$ is taken on the upper sheet such that $\lp{4}=0.23$ giving
\beq
 \int_1^4 \omega = \vect{0.0744 i}{0.230 i} ,
\label{a69} \eeq
and
\beq
\lambda = \lambda(\lp{4}) = C_\lambda\, \frac{\theta^2(a+\int_4^1)}{\hat{\theta}^2(a+\int_4^1)} = -0.597
\label{a70} \eeq
We can now write the vector $\zeta(\tau_+,\tau_-)$ that appears in the argument of the theta functions as (see eq.(\ref{a41}))
\beq
\zeta(\tau_+,\tau_-) = \vect{-0.295}{-0.403}  \tau_+ - \vect{0.0935}{0.349} \tau_-
\label{a71} \eeq
Finally, we need the constants
\beq
\mu_+ = - \tilde{\mu}_+ = 0.307\,i, \ \ \ \ \mu_- = -\tilde{\mu}_- = -1.765 \, i
\label{a72} \eeq
This allows us to use eqn.(\ref{a52b}) and write $\Psi(\lambda)$ as
\beq
\Psi(\lambda=-0.597) = \left(\begin{array}{cc} \re\, \psi_1 & \re\, \psi_2\\ \im\, \psi_1 & \im\, \psi_2\end{array}\right)
\label{a73} \eeq
In this case, the solutions $(\psi_1,\psi_2)$ and $(\tilde{\psi}_1,\tilde{\psi}_2)$ are complex and conjugate to each other. Therefore, we only need the real and imaginary part of one of them 
to have two independent solutions. The overall normalization $\det\Psi=1$ can be easily fixed later since $\det \Psi$ is a constant independent of $\tau_+,\tau_-$. 

\begin{figure}
    \centering
 \begin{subfigure}[b]{0.25\textwidth}
 \includegraphics[width=\textwidth]{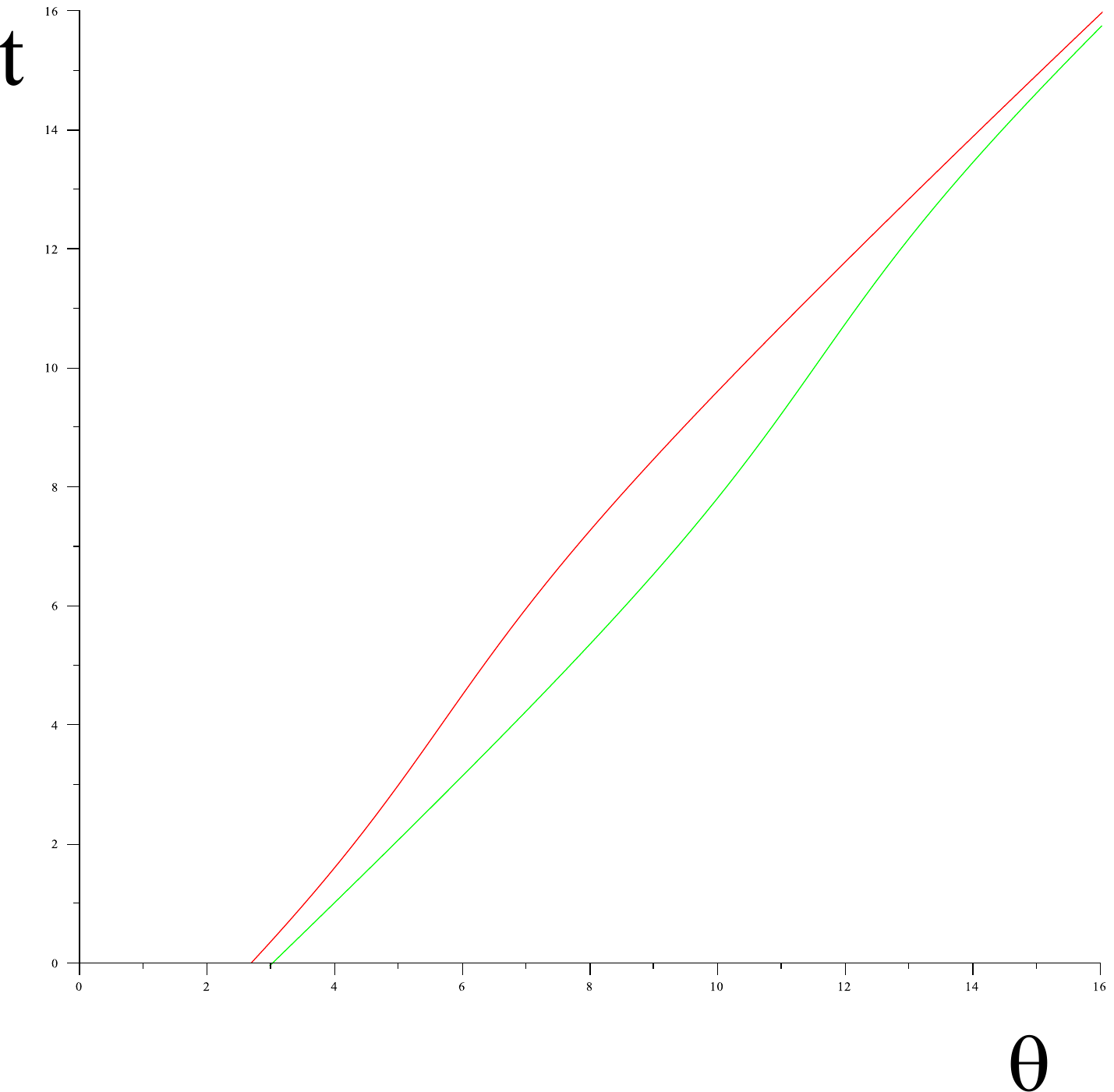}
 \end{subfigure}
 \begin{subfigure}[b]{0.3\textwidth}
 \includegraphics[width=\textwidth]{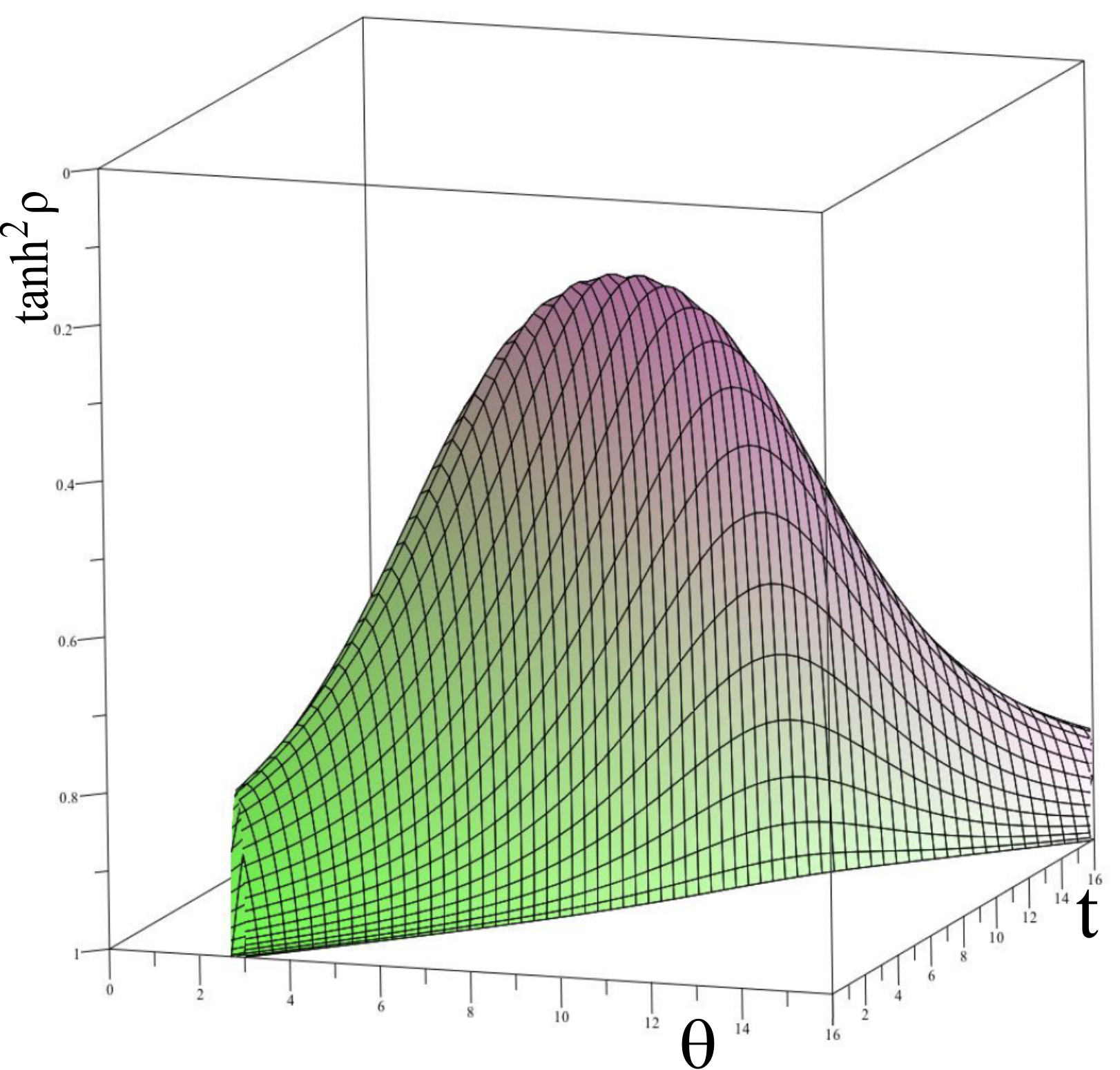}
 \end{subfigure}
 \begin{subfigure}[b]{0.3\textwidth}
 \includegraphics[width=\textwidth]{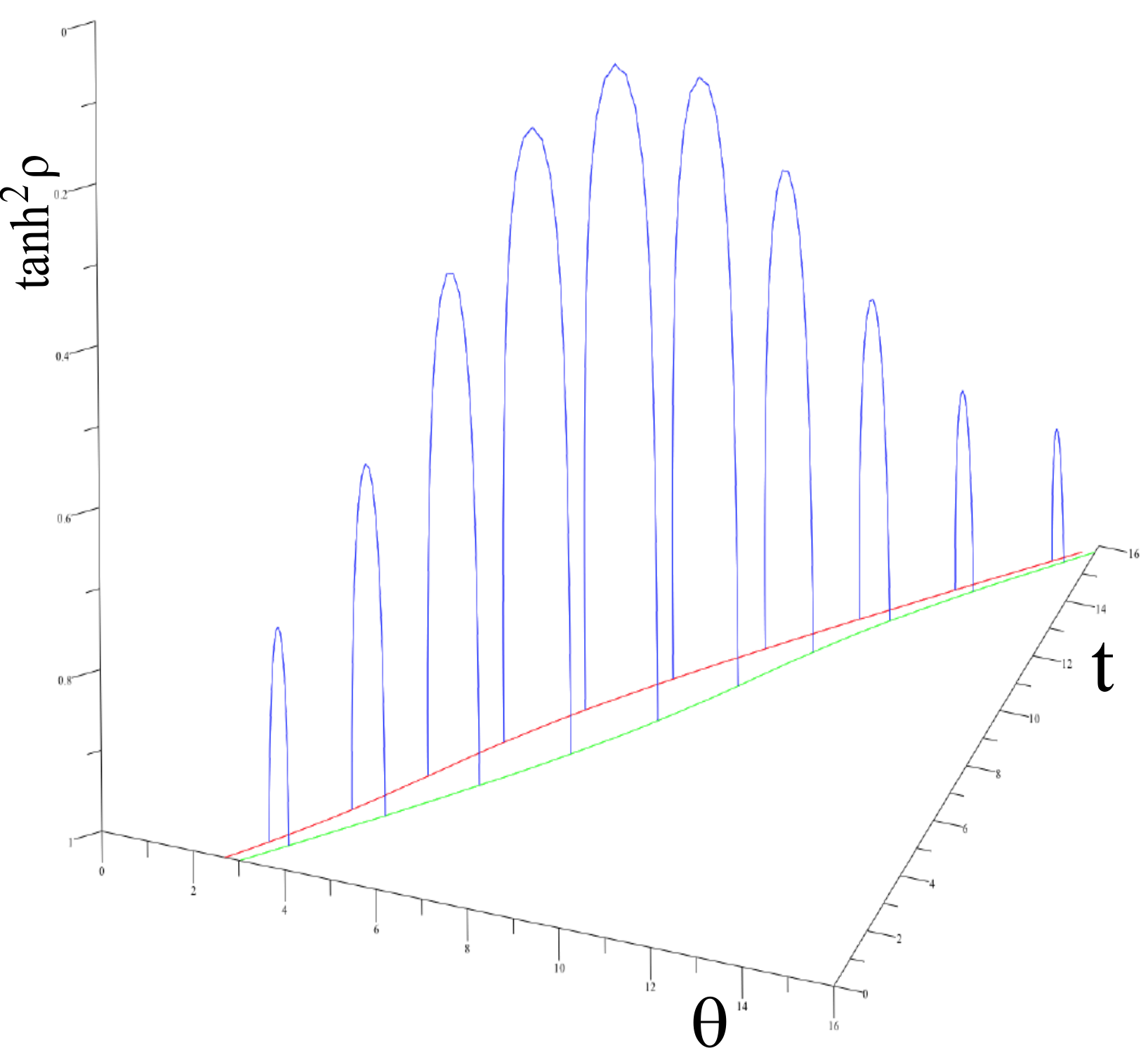}
 \end{subfigure}
  \caption{String ending on the boundary in global coordinates for the case of sinh-Gordon equation. The first picture displays the trajectories of the end points of the string in the boundary. The second and third one depicts the world-sheet in two different ways for clarity. The vertical axis is $\tanh^2\rho$. The bottom plane corresponds to $\rho=\infty$, \ie\ $\tanh^2\rho=1$.}
    \label{figsinh}
  \end{figure}

Finally, to construct $\mathbb{X}$, according to eq.(\ref{b14}), we also need to find $\Psi(-\lambda)$. For that purpose, choose
a new point $p'_4$ determined from equation eq.(\ref{a107}):
\beq
  \lp{4'} = -0.426\ .
\label{a74} \eeq
We are going to denote with a prime all quantities associated to this new choice of $p_4$. We find
\beq
  \int_1^{4'} \omega=\vect{-0.0372}{-0.383}\ ,
\label{a75} \eeq
and 
\beq
\lambda' = \lambda(\lp{4'}) = C_\lambda\, \frac{\theta^2(a+\int_{4'}^1)}{\hat{\theta}^2(a+\int_{4'}^1)} = .597 \ ;
\label{a76} \eeq
namely, indeed $\lambda'=-\lambda$. Finally, we have the new constants
\beq
 \mu'_+=-\tilde{\mu}'_+=-0.0868,\ \ \ \ \ \ \
  \mu'_-=- \tilde{\mu}'_-=-0.425 \ .
\label{a77} \eeq
 This allows us to put together $\Psi(\lambda'=-\lambda)$ using eqns.(\ref{a55}) with the new values of the constants:
\beq
\Psi(\lambda'=0.597) = \left(\begin{array}{cc} \psi'_1 & \psi'_2\\ \tilde{\psi}'_1 & \tilde{\psi}'_2\end{array}\right).
\label{a78} \eeq
In this case, the solutions $(\psi'_1,\psi'_2)$ and $(\tilde{\psi}'_1,\tilde{\psi}'_2)$ are real and linearly independent; we use them in constructing $\Psi(\lambda'=-\lambda)$. 
We now simply compute
\beq
\mathbb{X} = \Psi(\lambda) \Psi(-\lambda)^{-1} ,
\label{a79} \eeq
up to a normalization to ensure that $\det\mathbb{X}=1$. Notice that from the equations, $\det\mathbb{X}$ is independent of $(\tau_+,\tau_-)$ so this is just an overall normalization. 
The solutions are plotted in fig.\ref{figsinh}. The coordinates used are global coordinates $(t,\theta,\rho)$:
\beqa
 t &=& \arctan \frac{\X_{11}+\X_{22}}{\X_{12}-\X_{21}} ,\\
 \theta &=& \arctan \frac{\X_{11}-\X_{22}}{\X_{12}+\X_{21}}  ,\\
 \tanh^2 \rho &=& \frac{\tr(\X\X^t)-2\det\X}{\tr(\X\X^t)-2\det\X}  .
\label{a80} \eeqa
In the plots, for a better visualization, the coordinate $0<\tanh^2\rho <1$ was used instead of $\rho $. The horizontal plane $\tanh^2\rho=1$ is the boundary.
It is interesting to note that, near the boundary, the world-sheet is almost light-like and one can wonder if the near speed of light expansion discussed
in \cite{spinchain} may apply here. Although from the boundary point of view this is not clear, from the bulk point of view, the solutions discussed here and
the solutions discussed in \cite{JJ, DV} are essentially the same up to the reality conditions. In that sense an open string is somehow the analytic continuation 
of a closed string as already noted in \cite{AnCont}. We leave this as an interesting topic for further analysis and we discuss now an
example for the cosh-Gordon case.

 \subsection{cosh-Gordon}
\begin{figure}
   \centering
\begin{subfigure}[b]{0.25\textwidth}
\includegraphics[width=\textwidth]{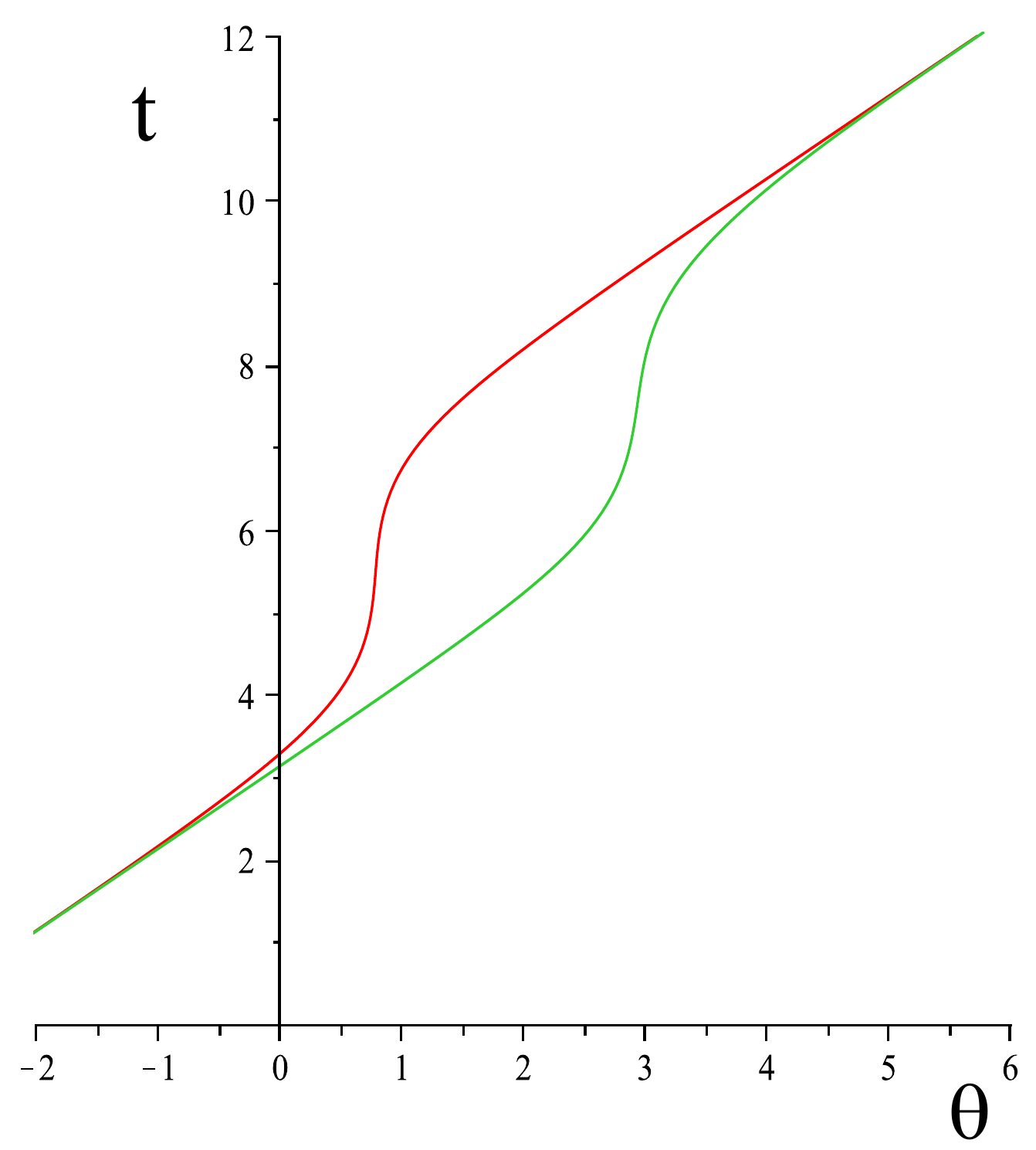}
\end{subfigure}
\begin{subfigure}[b]{0.3\textwidth}
\includegraphics[width=\textwidth]{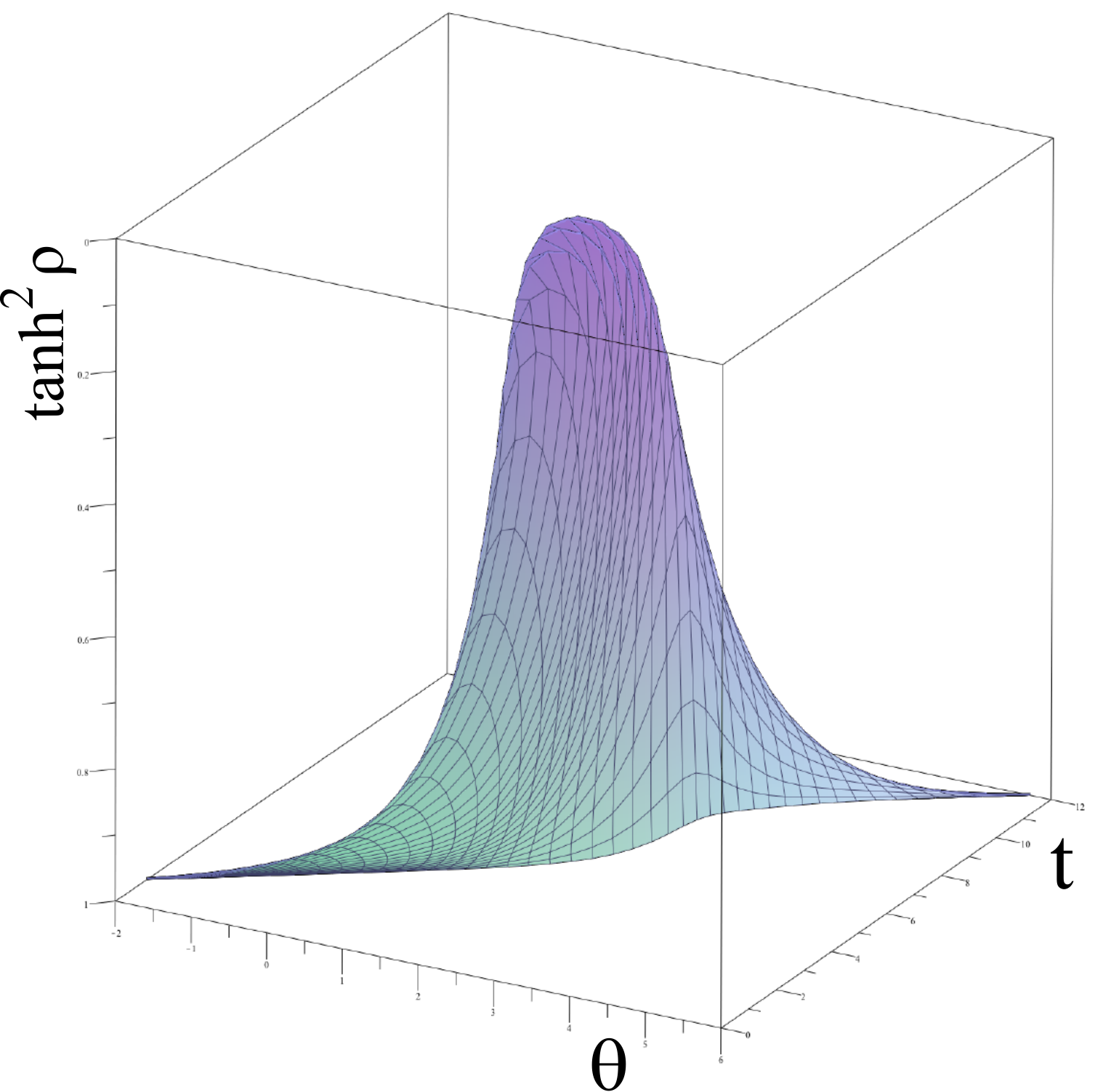}
\end{subfigure}
\begin{subfigure}[b]{0.3\textwidth}
\includegraphics[width=\textwidth]{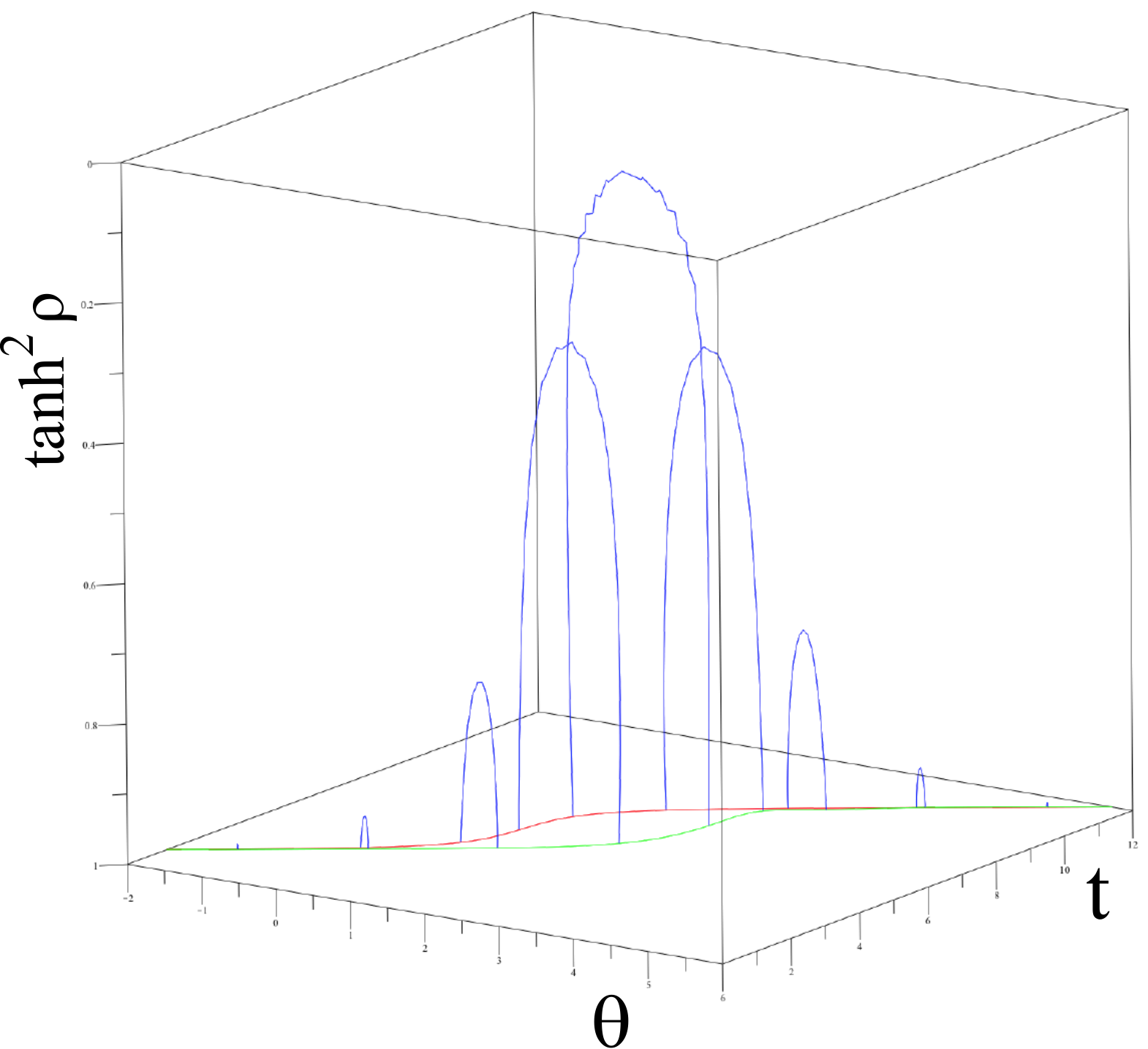}
\end{subfigure}
   \caption{String ending on the boundary in global coordinates for the case of cosh-Gordon equation. The first picture displays the trajectories of the end points of the string in the boundary. The second and third one depicts the world-sheet in two different ways for clarity. The vertical axis is $\tanh^2\rho$. The bottom plane corresponds to $\rho=\infty$, \ie\ $\tanh^2\rho=1$.}
   \label{figcosh}
 \end{figure}

 A similar example can be used to illustrate the cosh-Gordon case. The calculations are completely similar so we just go briefly over them.
 In fact, we can take the same Riemann surface but choose $p_1=0$, $p_3=-1$ which determines the characteristic of $\hat{\theta}$ to be
 \beq
  \Delta_1 = \vect{1}{1},\ \ \ \Delta_2 =\vect{0}{1} ,
 \label{a81} \eeq
 which is now odd and leads to the cosh-Gordon equation.
  The constants are
 \beq
 C_{\alpha}=-i,\ \  C_+=1,\ \  C_-=1,\ \  C_1=1,\ \  C_2=-i\ .
 \label{a82} \eeq
 We choose again $\lp{4}=0.23$ although this time we obtain $p'_4=-0.158$.  This determines
 \beq
  \int_1^4 \omega = \vect{0.0744 i}{0.230 i}, \ \ \ \  \int_1^{4'} \omega = \vect{-0.0449}{-0.203},
 \label{a83} \eeq
 and
 \beq
 \lambda = -\lambda' =\lambda(\lp{4}) = C_\lambda\, \frac{\theta^2(a+\int_4^1)}{\hat{\theta}^2(a+\int_4^1)} = -0.374 \ .
 \label{a84} \eeq
 The argument of the theta function is
 \beq
 \zeta(\tau_+,\tau_-) = \vect{-0.511 i}{0.187 i}  \tau_+ - \vect{0.0935 i}{0.349 i} \tau_-\ .
 \label{a85} \eeq
 Finally, we compute the constants
 \beq
 \mu_+ = - \tilde{\mu}_+ = -0.246, \ \ \ \ \mu_- = -\tilde{\mu}_- = -1.765, 
 \label{a86} \eeq
 \beq
 \mu'_+ = - \tilde{\mu}'_+ = -0.190\,i, \ \ \ \ \mu'_- = -\tilde{\mu}'_- = 1.484 \, i \ .
 \label{a87} \eeq
 We reconstruct the solution 
 \beq
 \Psi(\lambda=-0.374) = \left(\begin{array}{cc} \psi_1 & \psi_2\\ \tilde{\psi}_1 & \tilde{\psi}_2\end{array}\right), 
 \label{a88} \eeq
 \beq
  \Psi(\lambda=0.374) = \left(\begin{array}{cc} \re\, \psi'_1 & \re\, \psi'_2\\ \im\, \tilde{\psi}'_1 & \im\, \tilde{\psi}'_2\end{array}\right) .
  \label{a89} \eeq
The resulting $\mathbb{X}=\Psi(\lambda)\Psi^{-1}(-\lambda)$ is plotted in fig.\ref{figcosh}. 

\section{Acknowledgments}

We are grateful to Peter Ouyang for discussions and to A. Tseytlin for a discussion that lead to this paper and also comments on the final version.
This work was supported in part by NSF through grants PHY-0805948, a CAREER Award PHY-0952630, by DOE through grant DE-FG02-91ER40681 and by the SLOAN Foundation. In addition, M.K. 
is grateful to Perimeter Institute and M.K. and A.I. to the Simons Center for Geometry and Physics for hospitality 
while part of this work was being done.

 \appendix
 \section{Theta function identities}\label{sec7}
 Given a symmetric matrix $\Pi\in\mathbb{C}^{g\times g}$ with positive definite imaginary part, the theta function is defined as
 \beq
 \theta(\zeta) = \sum_{n\in\mathbb{Z}^g} e^{i\pi n^t \Pi n+2\pi i n^t\zeta},\ \ \ \ \ \zeta\in \mathbb{C}^g\ ,
 \label{a90} \eeq
 We also need another theta function defined as
 \beqa
 \hat{\theta}(\zeta) &=& \theta\left[\begin{array}{c} 
 \Delta_1 \\
 \Delta_2
 \end{array}\right] \\
 &=& 
 \exp\left\{2\pi i\left[\frac{1}{8}\Delta_1^t \Pi \Delta_1+\half \Delta_1^t \zeta +\frac{1}{4}\Delta_1^t \Delta_2 \right]\right\} 
           \theta\left(\zeta+\half\Delta_2+\half\Pi \Delta_1\right)\ , \nonumber
 \label{a90b} \eeqa 
 where $\Delta_{1,2}$ are vectors whose components are either zero or one. It is important to notice if $\Delta_1^t\Delta_2$ is even or odd since that determines the parity of $\hat{\theta}$:
 \beq
 \hat{\theta}(-\zeta) = e^{i\pi\Delta_1^t\Delta_2}\hat{\theta}(\zeta).
 \label{a91} \eeq
 Although the matrix $\Pi$ is an arbitrary complex matrix (up to the above mentioned conditions), when it is taken to be the period matrix of a hyperelliptic Riemann surface, the corresponding theta function has special properties.
 To be precise, consider the Riemann surface defined as a subspace of $\mathbb{C}^2$ (parameterized by $(\mu, \lambda)$  ) by the equation
 \beq
 \mu^2 = \prod_{i=1}^{2g+1} (\lambda-\lambda_i).
 \label{a92} \eeq
 As discussed in section \ref{sec5} below eq.(\ref{a38}), points on the Riemann surface are denoted as $p_i$ whereas their projection is $\lambda_{p_i}$. Define a basis of cycles $(a_i,b_i)$ and consider the differentials
 \beq
  \nu_i = \frac{\lambda^{i-1}}{\mu} d\lambda, \ \ \ \ i=1\cdots g \ .
 \label{a93} \eeq
 The index is restricted to $i=1\cdots g$ so that the differential is regular at zero and infinity. 
 We compute the $g\times g$ complex matrices
 \beq
  C_{ij} = \oint_{a_i} \nu_j , \ \ \ \ \tilde{C}_{ij}=\oint_{b_i} \nu_j ,
 \label{a94} \eeq
 which allows to choose a basis of normalized holomorphic differentials $\omega_i$ as
 \beq
 \omega_i =  \nu_j\,C^{-1}_{ji},
 \label{a95} \eeq
 and compute the period matrix
 \beq
 \Pi_{ij} = \oint_{b_i} \omega_j = \tilde{C}_{ik} C^{-1}_{kj}.
 \label{a96} \eeq
 When $\Pi$ is the period matrix of a Riemann surface, as defined above, the theta functions obey the trisecant identity\footnote{In perturbative 
 string theory the trisecant identity can be derived as a consequence of bosonization identities \cite{bosonization}.}:
 \beqa
 \lefteqn{\theta(\zeta)\; \theta\left(\zeta+\int_{p_{2}}^{p_1}\!\!\!\! \omega+\int_{p_3}^{p_{4}}\!\!\!\!\omega\right) =} \ \ \ \ && \label{eq:fay} \\
 &=& \gamma_{1234}\, \theta\Big(\zeta+\int_{p_{2}}^{p_{1}}\!\!\!\!\omega \Big)\, \theta\Big(\zeta+\int_{p_{3}}^{p_{4}}\!\!\!\!\omega \Big)
  +\gamma_{1324}\, \theta\Big(\zeta+\int_{p_{3}}^{p_{1}}\!\!\!\!\omega \Big) \,\theta\Big(\zeta+\int_{p_{2}}^{p_{4}}\!\!\!\!\omega \Big) \ , \nonumber
 \eeqa
where $p_i$ denote points on the Riemann surface and $\gamma_{ijkl}$ is defined as
\beq
\gamma_{ijkl}=\frac{\theta(a+\int_{p_{k}}^{p_{i}}\!\!\omega)\, \theta(a+\int_{p_{l}}^{p_{j}}\!\!\omega)}
      {\theta(a+\int_{p_{l}}^{p_{i}}\!\!\omega)\, \theta(a+\int_{p_{k}}^{p_{j}}\!\!\omega)} .
\label{a97} \eeq
Notice that $\int_{p_{j}}^{p_i}\!\! \omega$ denotes a complex vector in $\mathbb{C}^g$ since $\omega$ denotes the vector: $\omega= (\omega_1,\cdots,\omega_g)$.
Taking derivatives with respect to the position of the different points, we obtain a table of derivatives for the theta functions of interest. The most important ones are 
 \beq
  D_{13} \ln \left(\frac{\theta(\zeta)}{\hat{\theta}(\zeta)}\right)^2 = 2\frac{D_1\theta(a)D_3\theta(a)}{\hat{\theta}^2(a)} e^{i\pi\Delta_1^t\Delta_2}\left\{\left(\frac{\theta(\zeta)}{\hat{\theta}(\zeta)}\right)^2 - e^{i\pi\Delta_1^t\Delta_2}\left(\frac{\hat{\theta}(\zeta)}{\theta(\zeta)}\right)^2\right\},
  \label{a98} \eeq
  which relates to the sinh/cosh-Gordon equations and 
 \beqa
  D_3 \ln \frac{\theta(\zeta)}{\theta(\zeta+\int_1^4)} &=& 
  -D_3 \ln\frac{\hat{\theta}(a)}{\hat{\theta}(a+\int_4^1)} - \frac{D_3\theta(a)\theta(a+\int_4^1)}{\hat{\theta}(a)\hat{\theta}(a+\int_4^1)} \frac{\hat{\theta}(\zeta)\hat{\theta}(\zeta+\int_1^4)}{\theta(\zeta)\theta(\zeta+\int_1^4)}, \nonumber\\
  D_3 \ln \frac{\hat{\theta}(\zeta)}{\hat{\theta}(\zeta+\int_1^4)} &=& 
  -D_3 \ln\frac{\hat{\theta}(a)}{\hat{\theta}(a+\int_4^1)} -e^{i\pi\Delta_1^t\Delta_2} \frac{D_3\theta(a)\theta(a+\int_4^1)}{\hat{\theta}(a)\hat{\theta}(a+\int_4^1)} \frac{\theta(\zeta)\theta(\zeta+\int_1^4)}{\hat{\theta}(\zeta)\hat{\theta}(\zeta+\int_1^4)}, \nonumber \\
 D_1 \ln \frac{\theta(\zeta)}{\hat{\theta}(\zeta+\int_1^4)} &=& 
 -D_1 \ln\frac{\hat{\theta}(a)}{\theta(a+\int_4^1)} - \frac{D_1\theta(a)\hat{\theta}(a+\int_4^1)}{\hat{\theta}(a)\theta(a+\int_4^1)}  \frac{\hat{\theta}(\zeta)\theta(\zeta+\int_1^4)}{\theta(\zeta)\hat{\theta}(\zeta+\int_1^4)}, \nonumber\\ 
 D_1 \ln \frac{\hat{\theta}(\zeta)}{\theta(\zeta+\int_1^4)} &=& 
 -D_1 \ln\frac{\hat{\theta}(a)}{\theta(a+\int_4^1)} - \frac{D_1\theta(a)\hat{\theta}(a+\int_4^1)}{\hat{\theta}(a)\theta(a+\int_4^1)} \frac{\theta(\zeta)\hat{\theta}(\zeta+\int_1^4)}{\hat{\theta}(\zeta)\theta(\zeta+\int_1^4)}, 
 \label{a99} \eeqa
 which are used to solve for $\psi_{1,2}$. 
 The directional derivatives $D_{1,3}$ denote
 \beq
  D_{1,3} f(\zeta) = \lp{1,3}^{j-1} C^{-1}_{ji} \frac{\partial f(\zeta)}{\partial \zeta_i} .
 \label{a100} \eeq
 Notice that in all formulas, $D_{1,3}$ appear in such a way that they are defined up to a constant. The reason for defining this directional derivatives is that 
 \beq
 \frac{\partial}{\partial p_1} \int_{p}^{p_1} \omega = \omega_{p_1} .
 \label{a101} \eeq
 Since 
 \beq
 \omega_i = \frac{\lambda^{j-1}}{\mu(\lambda)} C_{ji}^{-1}, 
 \eeq
the directional derivative $D_1$ appears as a result of deriving with respect to the position of $p_1$. 
 The factor $\frac{1}{\mu(\lambda)}$ can be omitted since it is an overall factor.  Moreover, it is convenient to define the rescaled vectors
 \beq
 \hat{\omega}_1 = \frac{\hat{\theta}(a)}{D_1\theta(a)} \,\omega_1, \ \ \  \ \hat{\omega}_3 =  \frac{\hat{\theta}(a)}{D_3\theta(a)} \, \omega_3.
 \eeq
 In this way, defining 
 \beq
  \zeta = C_- \hat{\omega}_1 \tau_- + C_+ \hat{\omega}_3 \tau_+\, ,
 \eeq
 given any function of $\zeta$ we can identify
\beq
    \partial_- = C_- \frac{\hat{\theta}(a)}{D_1\theta(a)} \, D_1 ,\ \ \ \ \  \partial_+ = C_+ \frac{\hat{\theta}(a)}{D_3\theta(a)} \, D_3 .
    \label{a42} \eeq
Now it is straight-forward to check that eq.(\ref{a98}) reduces to the sinh/cosh-Gordon equation as stated in the main text.

\subsection{The spectral parameter}

In the main text, we found the following expression for the spectral parameter
\beq
 \lambda(\lp{4}) = C_\lambda \frac{\theta^2(a+\int_4^1)}{\hat{\theta}^2(a+\int_4^1)}
\label{a102} \eeq
where $a$ is an odd half-period and, therefore, a zero of the theta function. 
The purpose of this section is to simplify this expression. Although the procedure is well-known \cite{FK,ThF}, we reproduce the main steps since the result is of importance
to us. The first step is to study precisely where the function $\lambda(\lp{4})$ is defined. In the right hand side, a point $p_4$ on the Riemann surface
and a path connecting a fixed point $p_1$ to $p_4$ should be chosen. Different paths can only differ by a closed cycle $\mathcal{C}$ which in the basis $(a_i,b_i)$ can
be written as
\beq
 \mathcal{C} = \epsilon_{1i} b_i + \epsilon_{2i} a_i
 \label{a103} \eeq
 for some integer vectors $\epsilon_{1,2}$. Using the periodicity properties of the theta functions, it is straight forward to check that
 \beq
 \frac{\theta^2(a+\int_4^1+\epsilon_2 + \Pi \epsilon_1)}{\hat{\theta}^2(a+\int_4^1+\epsilon_2 + \Pi \epsilon_1)} = \frac{\theta^2(a+\int_4^1)}{\hat{\theta}^2(a+\int_4^1)} ,
 \label{a104} \eeq
and, therefore, the function $\lambda(\lp{4})$ is independent of the choice of path. Now we can wonder what happens is we choose $p_4$ in the upper or lower
sheet but with the same projection of the complex plane. This simply replaces $\int_1^4 \rightarrow -\int_1^4$. Using the property that $\theta^2(\zeta)$ and $\hat{\theta}^2(\zeta)$ are even functions, we find
\beq
 \frac{\theta^2(a-\int_4^1)}{\hat{\theta}^2(a-\int_4^1)}=\frac{\theta^2(-a+\int_4^1)}{\hat{\theta}^2(-a+\int_4^1)}=\frac{\theta^2(a+\int_4^1-2a)}{\hat{\theta}^2(a+\int_4^1-2a)}=\frac{\theta^2(a+\int_4^1)}{\hat{\theta}^2(a+\int_4^1)} ,
 \label{a105} \eeq
where in the last identity we used that $a$ is a half period and therefore $2a$ is a period which does not change the functions as already shown. 
Thus, $\lambda(\lp{4})$ is a function defined in the complex plane, that is, it has no cuts. Since $\theta$ has no essential singularities or poles,  
$\lambda(\lp{4})$ is a ratio of polynomials with zeros at the zeros of  $\theta^2(a+\int_4^1)$ and poles at the zeros of $\hat{\theta}^2(a+\int_4^1)$. 
By Riemann's theorem, the function $\theta(a+\int_4^1)$ (also $\hat{\theta}(a+\int_4^1)$ ) has $g$ zeros on the Riemann surface (as a function of $p_4$). 
For example, since $\theta(a)=0$, $\lp{4}=\lp{1}$ is a zero. On the other hand, since the characteristic 
$\left[\begin{array}{c}  \Delta_1 \\ \Delta_2 \end{array}\right]$ of $\hat{\theta}$ is such that $\half\Delta_2+\half\Pi\Delta_1=\int_1^3\omega$ we have that $\lp{4}=\lp{3}$ is a pole. The other possible zeros or poles are actually the other branch points. By considering each of them one can see that the other possible zeros and poles coincide and cancel each other, so the only actual zeros and poles are the ones we discussed. We find that
\beq
  \lambda(\lp{4}) = \frac{\theta^2(a+\int_4^1)}{\hat{\theta}^2(a+\int_4^1)} = A_0 \frac{\lp{4}-\lp{1}}{\lp{4}-\lp{3}} .
\label{a106} \eeq
As an aside, notice that since $\lp{1,3}$ are branch cuts, $\lp{1}$ is a double zero and $\lp{3}$ is a double pole if considered on the Riemann surface.
The constant $A_0$ can be easily evaluated by specializing $\lp{4}$ to a branch point other than $\lp{1,3}$, or by taking the limit of both sides
when $\lp{4}\rightarrow\lp{1,3}$. The last possibility is more useful but we refrain from doing it in generality since a value for $A_0$ is not needed. 
The result we really need is, given $p_4$, how to find a point $p'_4$ such that $\lambda(p'_4)=-\lambda(p_4)$. The reason being that the solution is written 
in terms of $\X = \Psi(\lambda)\Psi(-\lambda)^{-1}$. This problem is easily solved given the expression on the right hand side of eq.(\ref{a106}). The answer is
 \beq
 \lp{4'} = \frac{\lp{4}(\lp{1}+\lp{3})-2\lp{1}\lp{3}}{2\lp{4}-\lp{1}-\lp{3}},
 \label{a107} \eeq
and then $\lambda(\lp{\bar{4}}) = - \lambda(\lp{4})$. 

 \subsection{Reality conditions}
 
In order to find proper solutions to the string equations of motion we need to ensure that the theta functions involved are real or in some cases, purely imaginary. From the definition, we find
 \beq
  \overline{\theta}(\zeta) = \sum_{n\in\mathbb{Z}^g} e^{-i\pi n^t \bar{\Pi}n - 2\pi i n^t \bar{\zeta}}
 \label{a108} \eeq
which is unrelated to $\theta(\zeta)$ unless the matrix $\Pi$ satisfies some reality condition. If we impose the symmetry of the Riemann surface under the involution $\lambda\leftrightarrow \bar{\lambda}$ 
then we have
\beq
 \overline{\mu^2(\lambda)} = \mu^2(\bar{\lambda}) = \prod_{i=1}^{2g+1} (\bar{\lambda}-\lambda_i) .
 \label{a109} \eeq
If the cycles $a_i$ and $b_i$ are chosen such that
 \beq
  \bar{a}_i = T_{ij} a_j, \ \ \ \bar{b}_i = -T_{ij} b_j,
 \label{a109b} \eeq
with $T$ a symmetric matrix such that $T^2=1$ then 
 \beq
 \bar{\Pi} = - T \Pi T ,
 \label{a110} \eeq
and
 \beq
 \overline{\theta(\zeta)} = \theta(-T\bar{\zeta}).
\label{a111} \eeq
Since $\theta$ is an even function, it will be real for those values of $\zeta$ such that
\beq
 -T\bar{\zeta}= \pm \zeta, \ \ \ \Longrightarrow \ \ \ \theta(\zeta)\in \mathbb{R}.
 \label{a112} \eeq
In the simplest case, the cuts are chosen on the real axis and, with the canonical choice of ($a_i$, $b_i$), $T$ is just the identity matrix.
 To understand what happens with $\hat{\theta}$ consider a generic theta function with characteristics. It is easy to see that
  \beq
   \overline{\theta\left[\begin{array}{c}\Delta_1\\ \Delta_2\end{array}\right](\zeta)} = \theta\left[\begin{array}{c}T\Delta_1\\ -T\Delta_2\end{array}\right](-T\bar{\zeta}).
  \label{a113} \eeq
Assume now that $\Delta_{1,2}$ are chosen such that
  \beq
  T \Delta_1 = (-)^{\epsilon_1} \Delta_1, \ \ \ T \Delta_2 = - (-)^{\epsilon_2} \Delta_2,
  \label{a114} \eeq
and that 
\beq
   -T \bar{\zeta} = (-)^{\epsilon_\zeta} \zeta .
  \label{a115} \eeq
If we take into account that $\hat{\theta}=\theta\left[\begin{array}{c}\Delta_1\\ \Delta_2\end{array}\right](\zeta)$ is even or odd according to $\Delta^t_1 
  \Delta_2$ is even or odd then we find
  \beq
  \begin{array}{lcl}
   \Delta_1^t \Delta_2 \ \mbox{is even}:& \ \ & \theta, \hat{\theta} \ \ \mbox{are real} \\
   \Delta_1^t \Delta_2 \ \mbox{is odd}:&  & \theta\ \mbox{is real}, 
     (-)^{\epsilon_2+\epsilon_\zeta}=\left\{\begin{array}{rcl} 1, &\hat{\theta}& \mbox{is real}\\-1, &\hat{\theta}& \mbox{is purely imaginary}\end{array} \right. 
     \end{array}
  \label{a116} \eeq
   In other words, when $\Delta_1^t\Delta_2$ is odd then $\hat{\theta}$ is real or imaginary depending if $\zeta^t\Delta_2$ is real or imaginary.
  If $\Delta_1^t\Delta_2$ is even then $\hat{\theta}$ is real. Also, $\theta$ is always real. This, of course is valid when all the assumptions (\ref{a109}),(\ref{a109b}), (\ref{a110}), (\ref{a112}), (\ref{a114}) are true.

\end{document}